# Phase Binarization in Mutually Synchronized Bias Field-free Spin Hall Nano-oscillators for Reservoir Computing

Sourabh Manna[1], Rohit Medwal[2], John Rex Mohan[3], Yasuhiro Fukuma[3,4], Rajdeep Singh Rawat[1#]

[1]*Natural Sciences and Science Education, National Institute of Education, Nanyang Technological University, 637616, Singapore.*

[2]*Department of Physics, Indian Institute of Technology Kanpur, Uttar Pradesh, 208016, India.*

[3]*School of Computer Science and System Engineering, Kyushu Institute of Technology, Iizuka, 820-8502, Japan.*

[4]*Research Center for Neuromorphic AI hardware, Kyushu Institute of Technology, Kitakyushu 808-0916, Japan.*

\# Correspondence: rajdeep.rawat@nie.edu.sg

# Abstract

Mutually coupled spin Hall nano-oscillators (SHNO) can exhibit binarized phase state, offering pathways to realize Ising machines and efficient neuromorphic hardware. Conventionally, phase binarization is achieved in coupled SHNOs via injecting an external microwave at twice of the oscillator frequency in presence of a biasing magnetic field. However, this technology poses potential challenges of higher energy consumption and complex circuit design. Moreover, fabrication-induced mismatch in SHNO dimensions may hinder mutual synchronization. Addressing these challenges, we demonstrate purely DC current-driven mutual synchronization and phase binarization of two non-identical nanoconstriction SHNOs without biasing magnetic field and microwave injection. We thoroughly investigate these phenomena and underlying mechanisms using micromagnetic simulation. We further demonstrate the bias field-free synchronized SHNO pair efficiently performing a reservoir computing benchmark learning task: sin and square wave classification, utilizing current tunable phase binarization. Our results showcase promising magnetization dynamics of coupled bias field-free SHNOs for future computing applications.

# Introduction

Reservoir computing[1,2] (RC) is a promising computing paradigm that harnesses the nonlinear dynamics of complex systems for efficient information processing. RC comprises a time-dependent recurrent neural network (RNN) known as the "reservoir" and a time-invariant "readout" that connects the reservoir to the output. Unlike traditional neural networks, only the readout is trained in RC using linear regression that significantly reduces the training cost[3]. On-chip RC hardware can be designed by exploiting the rich nonlinear dynamics of coupled nano-oscillators[4,5]. Parametrically driven oscillator networks are already established to implement phase logic operation via subharmonic injection lock-in (SHIL) method in Boolean computing[6–8]. In such oscillators, two possible phase states are observed (phase binarization) which represent the binary "0" and "1" in phase logic. This phase binarization can modulate the coupling between the oscillators and enhance the nonlinearity in the dynamic variables[9]. Therefore, phase-binarized coupled nano-oscillators hold significant potential for RC hardware implementation.

Recently, ferromagnet (FM)/heavy metal (HM) bilayer-based spin Hall nano-oscillators (SHNO) have emerged as one of the top-tier CMOS compatible high frequency nano-oscillators[10] for realizing RC hardware, owing to their inherent nonlinear magnetization dynamics[11–14], miniature footprint[14–16], straightforward fabrication and low-power operation[17]. Additionally, efficient control of magnetization dynamics through bias-current, magnetic field, voltage controlled magnetic anisotropy and microwave injection lock-in make them suitable for neuromorphic hardware design[13,16,18–21]. Multiple nanoconstriction (NC) SHNOs can be mutually coupled through propagating spin wave, exchange interaction and magnetodipolar interaction, eliminating the need for electrical interconnects in SHNO arrays[22–26]. The current state-of-the-art demonstrates SHIL induced phase binarization in a two-dimensional NC SHNO array where the inter-SHNO coupling is primarily mediated by magnetodipolar interaction[27], paving a way towards realizing spin Hall Ising machine for efficiently solving computationally hard combinatorial optimization problems. Therefore, phase binarization in mutually coupled SHNOs can be a promising way for designing RC hardware as well. Nevertheless, practical implementation of this route poses major challenges. The reliance on external biasing magnetic field for typical SHNO operation and external microwave source for implementation of SHIL lead to complexity in circuit design process, increased circuit area and higher energy consumption. Hence, exploring phase binarization routes that exploit intrinsic magnetization dynamics in coupled SHNOs systems without external biasing magnetic field or microwave sources is imperative.

Recent studies on bias field-free single SHNO have demonstrated nontrivial auto-oscillation properties including in-plane to out-of-plane transition of auto-oscillation trajectory[28–30], reduction of threshold current[31] and tunable spiking behavior[29,32]. Investigating mutual synchronization of bias field-free NC SHNOs, therefore, can unveil intriguing magnetization dynamics for realization of efficient RC hardware. However, challenges arise due to fabrication-induced mismatch in nanoconstriction widths causing a difference in characteristic frequencies of individual SHNOs in an NC array. This frequency mismatch is detrimental for their mutual synchronization. This issue can be resolved by introducing a gradient in the biasing magnetic field[24] which is technically not allowed for bias field-free SHNOs . Therefore, exploring routes for mutual synchronization of bias field-free SHNOs with non-identical NC widths is crucial for practical applications.

In this report we demonstrate purely DC current-driven phase binarization in a mutually synchronized NC SHNO pair without any biasing magnetic field as well as external microwave injection. The SHNO pair is designed by defining two NCs with non-identical widths, representing two different SHNOs separated by a distance ($d$) as shown in Fig. 1a. We comprehensively investigate the mechanism of mutual synchronization and DC current-driven phase binarization in bias field-free condition. More importantly, we demonstrate that the phase binarized bias field-free SHNO pair can efficiently perform an RC benchmark learning task: sin and square wave classification.

In our previous work, we demonstrated that an in-plane uniaxial anisotropy can induce bias field-free auto-oscillation of magnetization in a NiFe/Pt bilayer-based NC SHNO device[31]. Such a magnetic anisotropy can be induced in NiFe by implementing suitable growth schemes[33–36]. Hence, we chose a similar NiFe/Pt bilayer-based NC pair as shown in Fig. 1a. We defined the in-plane uniaxial anisotropy in the NiFe layer with the easy-axis oriented at an angle $\phi$ measured from the $x$-axis (Fig. 1a). The uniaxial anisotropy gives rise to the anisotropy field given as, $\boldsymbol{H}_{anis} = \frac{2K_u}{M_s}(\hat{\boldsymbol{u}} \cdot \boldsymbol{m})\hat{\boldsymbol{u}}$ , where $K_u$, $M_s$, $\boldsymbol{m}$ and $\hat{\boldsymbol{u}}$ are the uniaxial anisotropy constant, saturation magnetization of NiFe, reduced magnetization vector ($\boldsymbol{m} = \boldsymbol{M}/M_s$) and the unit vector along the easy axis respectively. In the stable equilibrium, the orientation of magnetization is determined by the cumulative effect of magnetic anisotropy, magnetostatic field and exchange interaction. As we pass a charge current through the SHNO device, it induces a transverse spin current via spin Hall effect in Pt. The spin current exerts spin-orbit torque (SOT) on the magnetization in NiFe and destabilizes it from the stable equilibrium. The SOT may enhance or oppose the intrinsic damping torque in the ferromagnet based on the mutual orientation of the magnetization and the spin polarization ($\boldsymbol{\sigma}$) of the injected spin current. In the latter case, a large enough SOT can completely compensate the intrinsic damping torque and lead to the limit cycle auto-oscillation of magnetization about the effective internal field ($\boldsymbol{H}_{eff}$). In absence of any biasing magnetic field, $\boldsymbol{H}_{eff}$ is given as:

$$\boldsymbol{H}_{eff} = \boldsymbol{H}_{anis} + \boldsymbol{H}_{demag} + \boldsymbol{H}_{exch} + \boldsymbol{H}_{Oe}, \qquad (1)$$

where $\boldsymbol{H}_{demag}$, $\boldsymbol{H}_{exch}$ and $\boldsymbol{H}_{Oe}$ denote the magnetostatic field, the exchange field and the DC current induced Oersted field respectively. The SOT driven magnetization dynamics can be quantitatively formulated in terms of the LLG equation with additional SOT term as follows[37,38].

$$\dot{\boldsymbol{m}} = -\gamma \boldsymbol{m} \times \boldsymbol{H}_{eff} + \alpha \boldsymbol{m} \times \dot{\boldsymbol{m}} + \frac{\gamma |J_c| \hbar \theta_{SH}}{2 e t_{FM} \mu_0 M_s} \boldsymbol{m} \times (\boldsymbol{\sigma} \times \boldsymbol{m}). \qquad (2)$$

In equation 2, $\alpha$ denotes the intrinsic damping parameter, $\theta_{SH}$ is the spin Hall angle of Pt which is taken as 0.08[19,22,23] in the present study. The quantities in the SOT term, $\gamma$, $\hbar$, $e$, $\mu_0$, $\boldsymbol{\sigma}$ and $t_{FM}$ represent the gyromagnetic ratio, reduced Planck's constant, electronic charge, permeability of vacuum, spin-polarization direction, and thickness of the NiFe layer respectively. Notably, the field-like component of SOT arising from the bulk of Pt and interfacial Rashba effect has been found to be significantly smaller than the damping-like component of SOT in NiFe/Pt system[38,39]. Therefore, the SOT driven magnetization dynamics in such system is efficiently modelled considering only the damping-like component of SOT. We have employed the micromagnetic modelling approach to numerically solve equation 2 (see Method) and obtained the oscillatory magnetization dynamics in bias field-free condition.

## Results

### Device design and auto-oscillation properties

We consider a NC SHNO pair consisting of a 5 nm thick NiFe layer possessing uniaxial anisotropy, interfaced with a 5 nm thick Pt layer (Fig. 1a). The two NCs of widths 100 nm (left NC) and 150 nm (right NC) are separated by a center-to-center distance, $d$ = 200 nm. From here on, we will refer to this NC SHNO pair as "NC pair". Figure 1a schematically represents the device structure of the NC pair and the spatial distribution of current density ($J_c$) in the Pt layer for 1 mA input current ($I_{DC}$), as obtained from Multiphysics simulation using COMSOL (see Methods). We explicitly consider the dominant $x$-component of $J_c$ for computing the current-induced spin-orbit torque as the $y$ and $z$-components are order of magnitude smaller than the $x$-component (see supplementary information, Fig. S1). Therefore, we defined $\sigma = -\hat{y}$ considering the orthogonality of $J_c$, $\sigma$ and the spin current ($J_s$) along $-\hat{z}$. As obvious from Fig. 1(a), $J_c$ at the left NC is higher as compared to the right NC. This essentially leads to lower threshold current for auto-oscillation in the left NC.

We numerically solved equation 2 using the GPU accelerated micromagnetic solver MuMax3[40] to obtain the SOT driven magnetization dynamics in the NC pair (see Methods for details of micromagnetic simulation). We explicitly defined $K_u$ = 7.5 kJ/m³ to account for the uniaxial anisotropy[31]. The solver computes $m_x$ as a function of time ($t$) at each micromagnetic cell. In the auto-oscillation state, the frequency of limit cycle oscillation has been extracted by performing fast Fourier transform (FFT) of the spatial average of $m_x(t)$ obtained from the entire geometry. In experiments, the oscillation of $m_x(t)$ gives rise to oscillating anisotropic magnetoresistance (AMR) which is coupled with $I_{DC}$ and subsequently converted into a longitudinal microwave voltage across the SHNO. The $m_y(t)$ and $m_z(t)$ are not associated with this AMR based detection technique. Therefore, we analyze the $m_x(t)$ to find out the auto-oscillation characteristics.

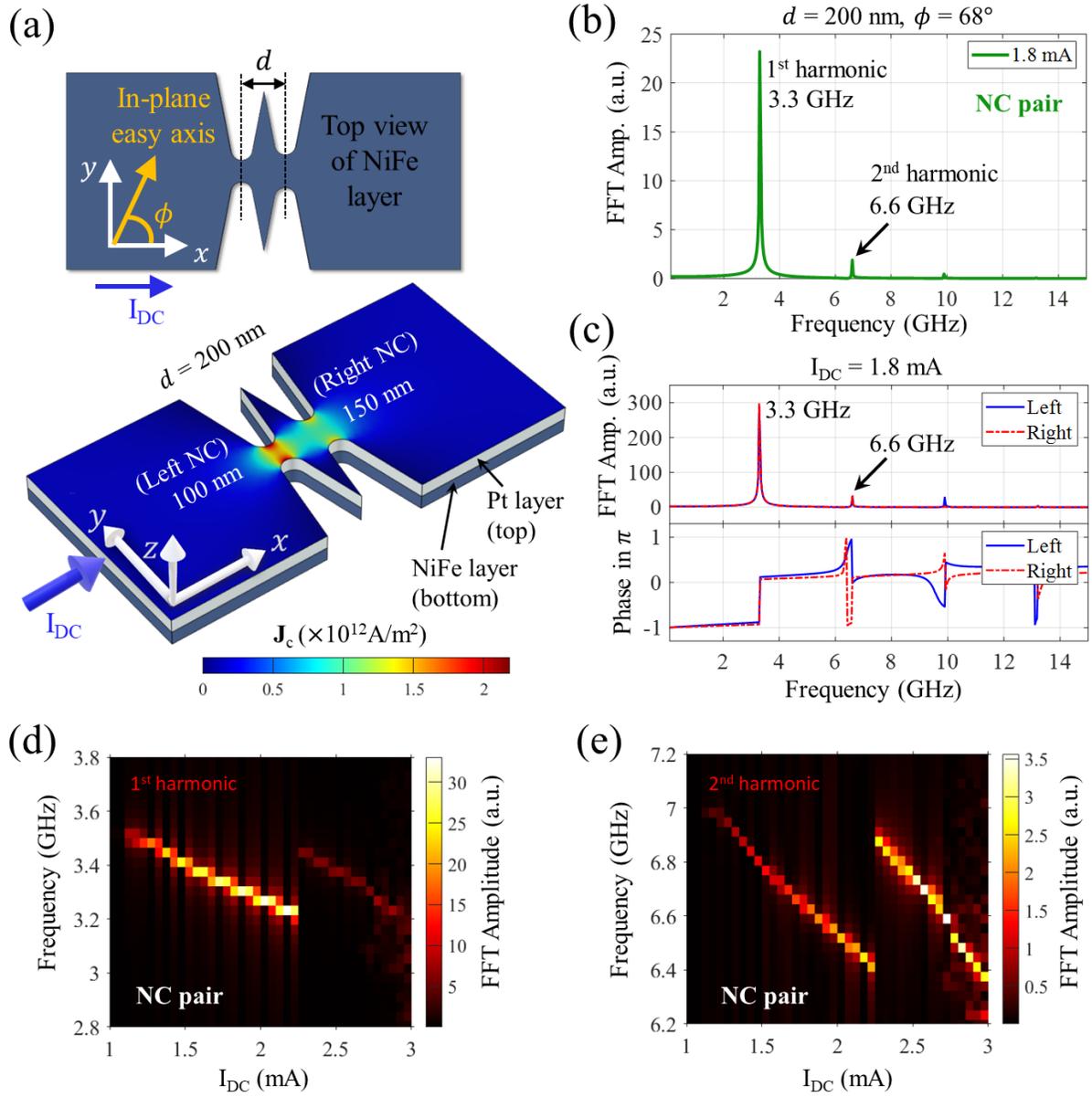

**Figure 1. | Schematic of the device geometry and auto-oscillation characteristics.** (a) The NC pair geometry designed in COMSOL and the spatial distribution of current density in the Pt layer. (b) FFT Spectrum of the auto-oscillation obtained from the entire NC pair geometry for $I_{DC}$ = 1.8 mA. (c) FFT spectra of the local auto-oscillation at individual NCs in the NC pair. (d) Current tunability of the 1st harmonic of auto-oscillation in the NC pair. (e) Current tunability of the 2nd harmonic of the auto-oscillation in the NC pair.

A typical FFT spectrum corresponding to the limit cycle auto-oscillation for $I_{DC}$ = 1.8 mA is shown in Fig. 1b. The sharp peak at 3.3 GHz is accompanied by a weak 2nd harmonic of 6.6 GHz. We further obtained the local auto-oscillation characteristics of individual NC SHNO within the NC pair by averaging $m_x(t)$ over 170nm $\times$ 170 nm area around the center of each NC and applying FFT. Figure 1c shows the frequency and phase extracted from the FFT of the local magnetization auto-oscillation profile in individual NCs. It is clear from Fig. 1c that both SHNOs in the NC pair exhibit the same auto-oscillation frequency and phase at 1.8 mA current, denoting mutual synchronization. Therefore, the auto-oscillation frequency of individual NC SHNOs (Fig. 1c) match with the auto-oscillation frequency of the entire NC pair (Fig. 1b). We further observe the current-tunability of the auto-oscillation frequency and amplitude, obtained from the entire NC pair for both 1st harmonic (Fig. 1d) and 2nd harmonic (Fig. 1e). The red shift behavior of the frequency as a function of $I_{DC}$ is consistent with our previous study on bias field-free single NC SHNO[31]. We notice from Fig. 1d and 1e that, despite the difference in the NC widths, the NC pair exhibits auto-oscillation synchronously for each $I_{DC}$ value at 1st harmonic as well as 2nd harmonic. This persistent mutual synchronization of both NC SHNOs for more than 1.5 mA span of $I_{DC}$ is comparable with the state-of-the-art in-plane field assisted SHNO pair[23]. For both harmonics, the NC array exhibits red shift in auto-oscillation with increase in FFT amplitude for $I_{DC}$ from about 1.1 to 2.2 mA. A sudden nonlinear blue shift is observed around 2.2 mA for both harmonics, which is explained later. For $I_{DC} \geq 2.25$ mA, both harmonics show the red shift again with increasing $I_{DC}$. However, while the FFT amplitude of 1st harmonic decreases significantly in this $I_{DC}$ regime, the amplitude of 2nd harmonic continues to increase.

To gain a deeper insight into the mutual synchronization phenomenon, we separately simulated and analyzed the bias field-free auto-oscillation in single NC SHNOs with 100 nm and 150 nm constriction widths. Figure 2a shows the comparison of the current tunability of characteristic auto-oscillation frequency (the dominant 1st harmonic) for 100 nm single NC SHNO (blue), 150 nm single NC SHNO (red) and the NC pair (green). We observe that for lower value of $I_{DC}$ (< 1.45 mA) the auto-oscillation in the NC pair is purely driven by the auto-oscillation of 100 nm wide NC (left NC). Therefore, the auto-oscillation frequency of the NC pair overlaps with the auto-oscillation frequency of 100 nm single NC SHNO. However, at $I_{DC} \geq 1.45$ mA, the magnetization auto-oscillation takes place at the 150 nm wide NC (right NC) as well. At this point both SHNOs couple with each other. Therefore, the auto-oscillation frequency of the NC pair deviates from the characteristic auto-oscillation frequencies of single NC SHNOs. This coupling is identified as negative coupling[9,41] as the auto-oscillation frequency in mutually synchronized state is lower than the individual frequencies of single NC SHNOs. However, above 2.2 mA, the auto-oscillation frequency at mutually synchronized state rapidly shifts towards higher frequency, as compared to the characteristic frequencies of single NC SHNOs. This indicates a positive coupling[41] between the SHNOs at higher current.

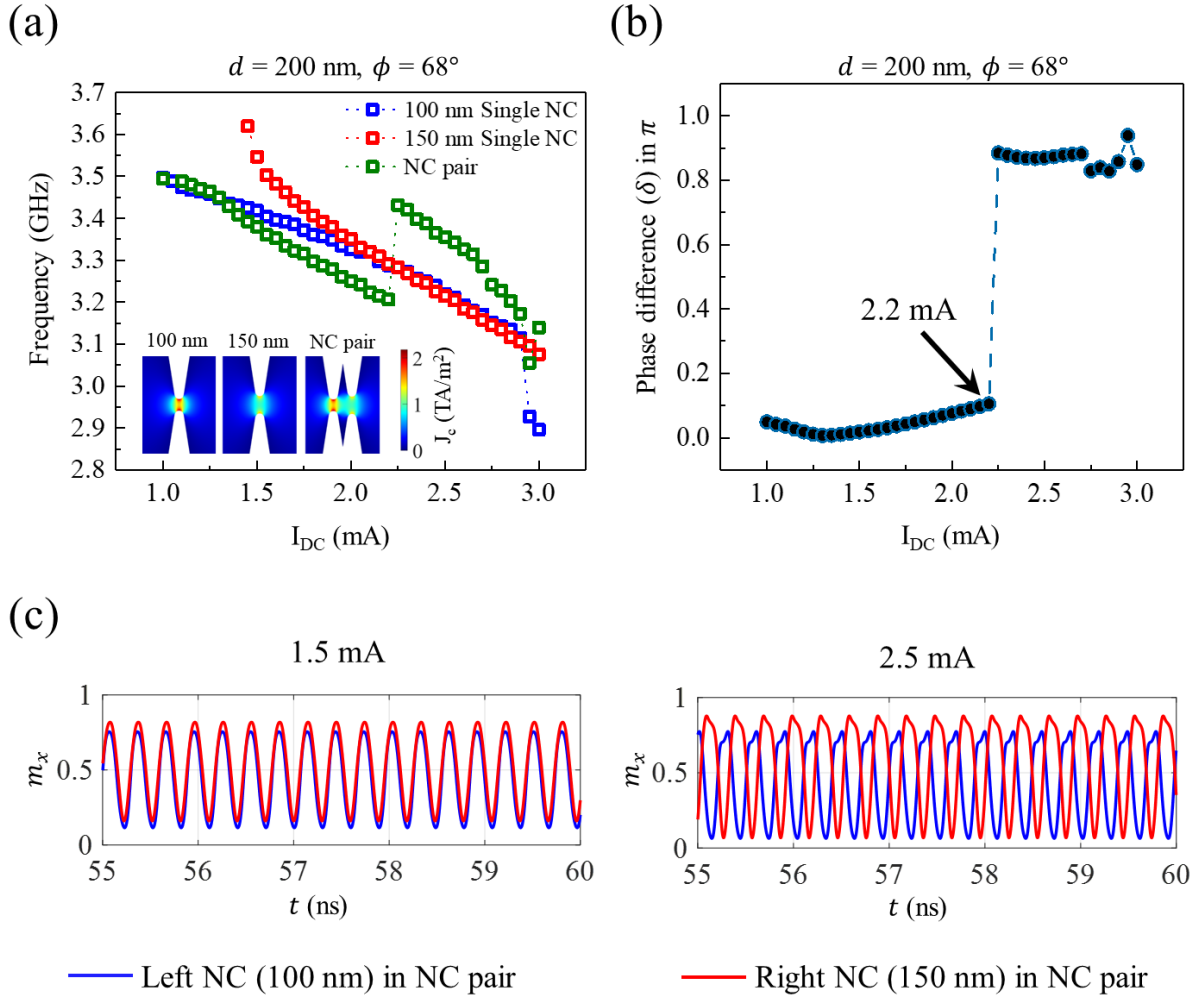

**Figure 2 | Mutual synchronization of NC SHNOs and phase binarization**. (a) Comparison of the current-tunability of auto-oscillation frequency in the NC pair at mutually synchronized state, with the singel NC SHNOs. The inset figure shows the current density distribution at the vicinity of the NCs in single NC SHNO and NC pair. (b) Phase difference for the 1st harmonic between the two NC SHNOs in the NC pair as a function of $I_{DC}$. The phase binarization threshold current 2.2 mA has been denoted by the pair. (c) Temporal profile of the local magnetization auto-oscillation in the individual NCs for input current of 1.5 mA (in-phase auto-oscillation) and 2.5 mA (out-of-phase auto-oscillation).

It should be noted that the frequency of auto-oscillation in such anisotropy-assisted bias field-free SHNO is primarily determined by the anisotropy field $H_{anis} = \frac{2K_u}{M_s}(\hat{u} \cdot m)\hat{u}$ as discussed in our previous report[31]. At higher input current, stronger SOT leads to increase in the amplitude of magnetization precession. Hence, $m$ moves away from the easy axis (along $\hat{u}$) resulting in a reduction of $H_{anis}$, which in turn reduces the auto-oscillation frequency. This explains the red-shift behavior of the auto-oscillation frequency in such bias field-free SHNOs. Therefore, the sudden blue-shift at $I_{DC} > 2.2$ mA can be attributed to the reduction of auto-oscillation amplitude which is also evident from the diminishing FFT amplitude in Fig. 1d. This could be possible if there is a substantial phase difference between the local magnetization auto-oscillation at the NCs. Despite the high amplitudes of local magnetization auto-oscillation at the NCs at higher $I_{DC}$, this phase difference would lead to a reduction of the resultant oscillation amplitude of $m$ in the NC pair. Hence, the NC pair exhibits a non-monotonic frequency variation as a function of $I_{DC}$ unlike the single NC SHNOs.

**Phase binarization in mutually synchronized state**

Now we proceed to investigate the behavior of the phase difference ($\delta$) for the 1$^{st}$ harmonic between the two SHNOs in the NC pair at mutually synchronized state. Figure 2b shows the behavior of $\delta$ as a function of $I_{DC}$. We notice that up to 2.2 mA, $\delta$ is quite small (within $\pi/9$ rad). That indicates both the NC SHNOs exhibit (nearly) in-phase auto-oscillation at this range of $I_{DC}$. In contrast, for $I_{DC} \geq 2.25$ mA current, $\delta$ becomes significant ($\delta \sim 8\pi/9$ rad) indicating (nearly) out-of-phase auto-oscillation in both NCs. Figure 2c shows the local $m_x(t)$ in both the NCs exhibiting in-phase auto-oscillation at $I_{DC} = 1.5$ mA and out-of-phase auto-oscillation at $I_{DC} = 2.5$ mA.

We observe that the NC pair exhibits two discrete auto-oscillation states in terms of phase difference ($\delta$) between the local magnetization auto-oscillation at both NCs. In the out-of-phase auto-oscillation state, the discretization of individual phases of the two NCs to two distinctive phase states is known as the "phase binarization"[27,42]. Of course, the phase binarization observed in our NC pair is not exactly the ideal case i.e., $\delta = \pi$. The reason will be explained later in this section. However, the phase difference between the individual NCs is large enough to classify the individual phases as *binarized*. Typically, the phase binarization in coupled oscillator system is realized through SHIL method[42], where the oscillation frequency ($f$) of the system is "locked" by an external periodic signal (referred as the "locking signal" later on) of frequency $2f$ [27]. We emphasize that the phase binarization in our NC pair has been achieved in absence of any biasing magnetic field and external locking signal, therefore it is purely driven by the input DC current.

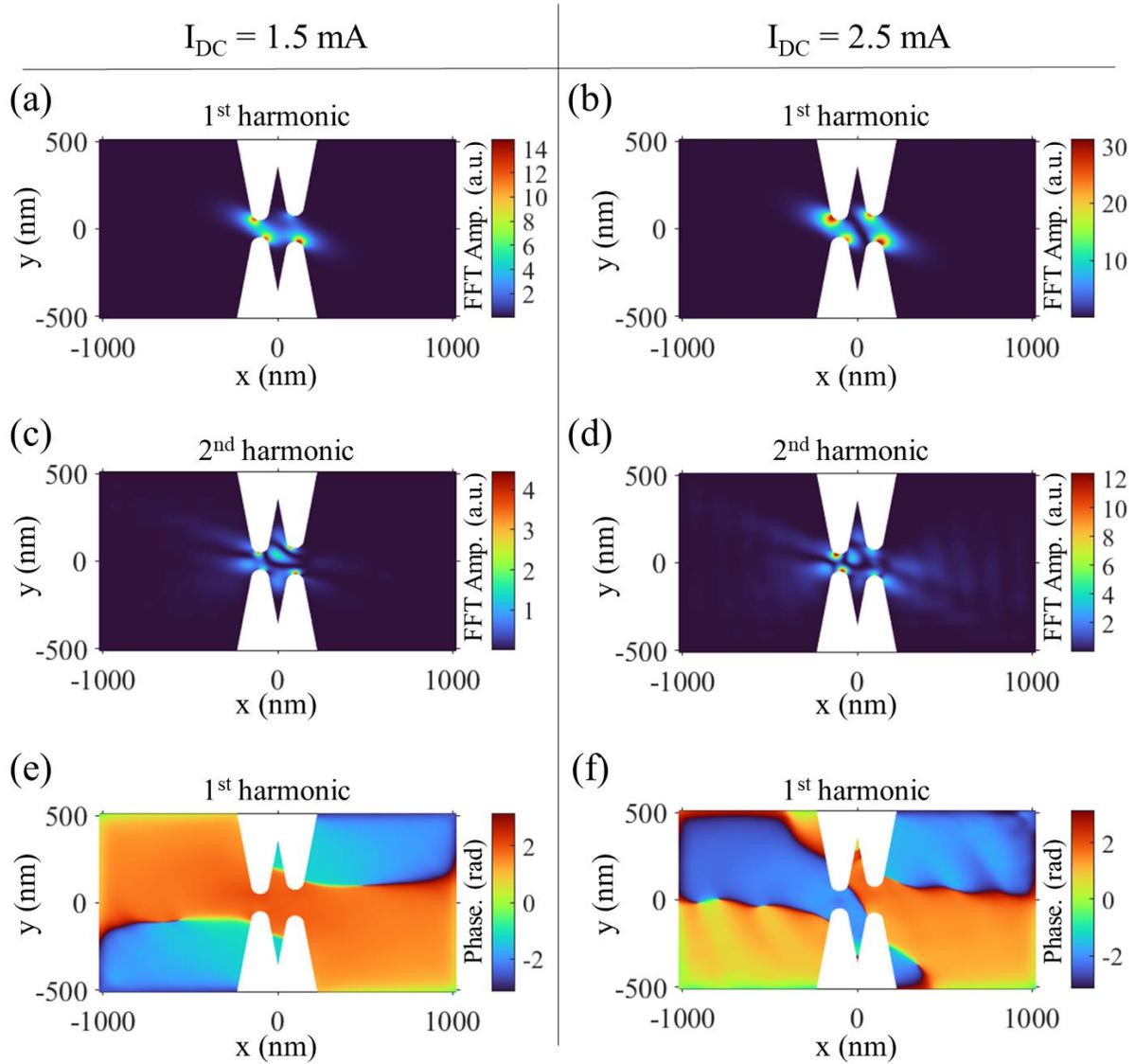

**Figure 3 | Spatial profile of auto-oscillation amplitude and phase in the NC pair.** The left colum shows a typical in-phase auto-oscillation ($I_{DC}$ = 1.5 mA) and the right column shows a typical phase-binarized auto-oscillation state ($I_{DC}$ = 2.5 mA). The spatial profiles of auto-oscillation amplitude have been shown for 1st harmonic in (a) and (b), and for 2nd harmonic in (c) and (d). The spatial profiles of phase at the 1st harmonic are shown in (e) and (f).

To understand the underlying mechanism of phase binarization in the NC pair, we first look at the spatial profile of auto-oscillation amplitude. We look for the interaction between spin wave modes particularly in the vicinity of the constrictions. Figure 3 depicts these spatial profiles for an in-phase ($I_{DC}$ = 1.5 mA) and a phase-binarized ($I_{DC}$ = 2.5 mA) auto-oscillation states. In both cases, the 1$^{st}$ harmonics are localized spin wave edge modes[19]. These edge modes overlap between the NCs for $I_{DC}$ = 1.5 mA (Fig. 3a), reinforcing the in-phase synchronization of the local auto-oscillations in both NCs. On the other hand, the edge modes of the two NCs are discrete in space for $I_{DC}$ = 2.5 mA (Fig. 3b) as observed by well demarcated existence of nearly zero FFT amplitude region in-between the two NCs. Therefore, in this case, the 1$^{st}$ harmonics (the dominant harmonics) of the auto-oscillation in both NCs hardly drive each other for in-phase synchronization. In contrast to the 1$^{st}$ harmonic, the 2$^{nd}$ harmonic is a propagating spin wave mode that creates the interference-like pattern in the spatial profile of auto-oscillation amplitude as seen in Fig. 3c and 3d (see supplementary information, Fig. S2 as well). Hence, the phase binarization in the 1$^{st}$ harmonic is possibly being mediated by the 2$^{nd}$ harmonic through SHIL mechanism, where one NC serves as the source of the *locking* periodic signal for the other NC. This is possible because of the characteristic frequency matching of the individual single NC SHNOs at higher input current (see Fig. 2a).

We confirm the 2$^{nd}$ harmonic mediated phase-binarization in our NC pair SHNO by analyzing the magnetization auto-oscillation obtained from a similar NC pair with $d$ = 800 nm. Figure 4 summarizes these results. In Fig. 4a we observe that the two NC SHNOs exhibit auto-oscillation at their free-running frequencies at lower currents unlike the $d$ = 200 nm NC pair. For $I_{DC}$ ~ 2.1 mA onwards they mutually synchronize as their characteristic frequencies match in that range (see Fig. 4a and 1a). The phase binarization (Fig. 4c) and its effect on auto-oscillation amplitude (Fig. 4b) have been clearly observed in the synchronized auto-oscillation state. Note that the out-of-phase state is quite close to the ideal case of phase binarization i.e., $\delta$ ~ $\pi$ rad. However, the other phase state corresponds to $\delta$ ~ $2\pi/5$ rad is *weakly binarized*. The difficulty of in-phase synchronization is due to the longer separation between the NCs resulting in no overlapping of the localized edge modes as shown in Fig. 4d. In contrast to the localized 1$^{st}$ harmonic, the 2$^{nd}$ harmonic spin wave modes propagate from one NC to the other, resulting in generation of interference-like pattern as shown in Fig. 4e. Therefore, the propagating 2$^{nd}$ harmonic reinforces the out-of-phase auto-oscillation (see the spatial phase map in Fig. 4f) through SHIL.

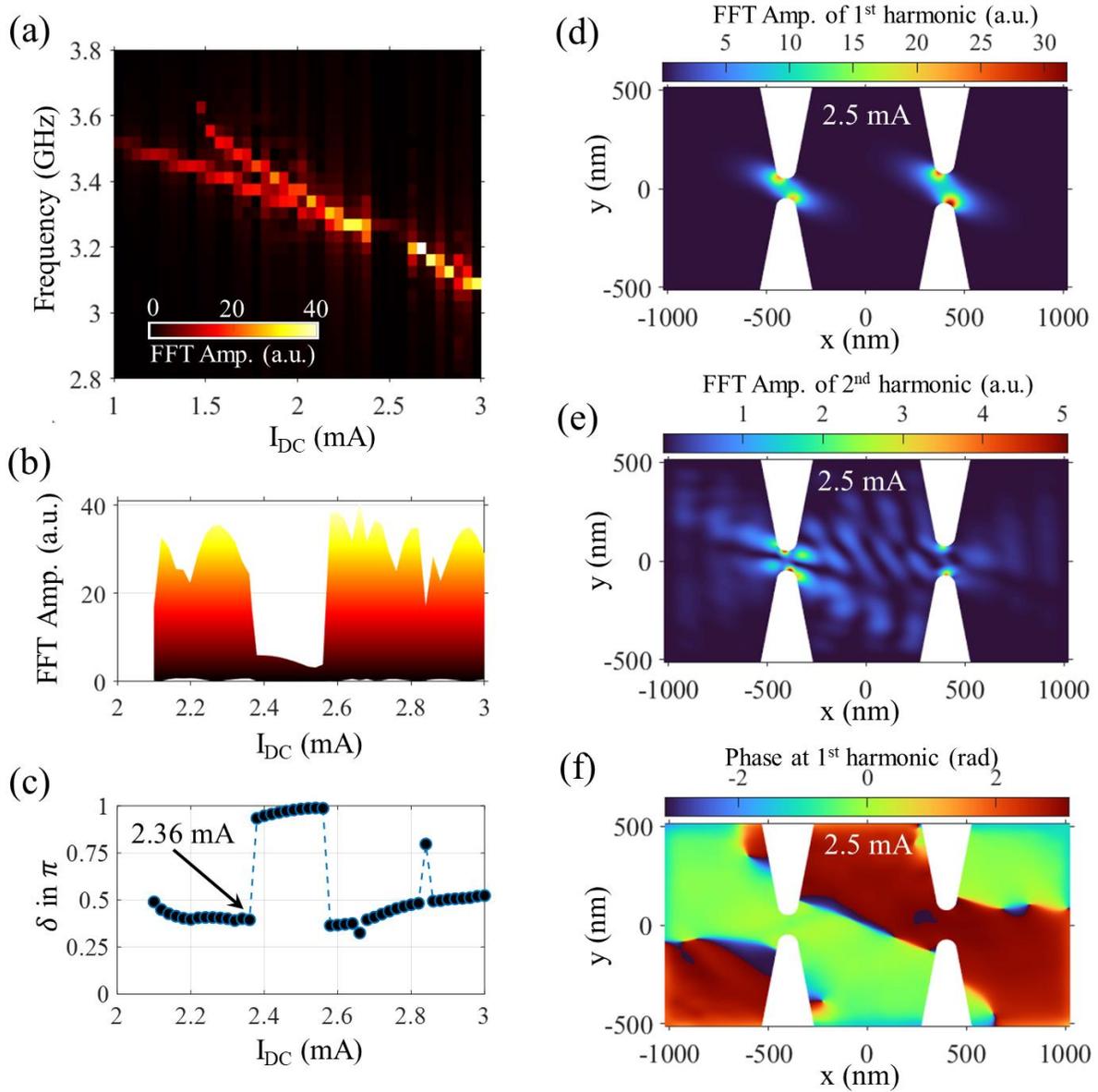

**Figure 4 | Auto-oscillation characteristics in NC pair with $d$ = 800 nm**. (a) Current-tunability of auto-oscillation frequency obtained from the entire NC pair geometry. (b) Variation of FFT amplitude as a function of $I_{DC}$ in the mutually synchronized state. (c) Phase difference between the local auto-oscillation in the NCs. The phase binarization threshold current is denoted as 2.36 mA. (d), (e) Spatial profile of auto-oscillation amplitude at 1st and 2nd harmonic respectively in phase binarized state. (f) Spatial profile of phase of 1st harmonic at the phase bianrized state.

We further note that the spin waves exhibit amplitude decay as they propagate due to the damping present in the ferromagnet. When the 2nd harmonic spin wave travels over 800 nm distance between the NCs, it experiences relatively higher decay in intensity as compared to the previous case of 200 nm separation between the NCs. Consequently, for identical input current, the SHIL induced phase binarization takes place at higher threshold current. As seen from Fig. 2b and Fig. 4c, the phase binarization threshold is 2.2 mA for $d$ = 200 nm NC array, and 2.36 mA for $d$ = 800 nm NC array.

Lastly, we would like to mention that the SHIL induced phase binarization strongly depends on the power of the locking signal. SHIL induced stable phase binarization in bias field assisted SHNO array is achieved above a certain threshold power of the locking signal. However, these SHNOs may switch between in-phase and out-of-phase auto-oscillation states below the threshold power of the locking microwave[27]. In the present bias field-free SHNO pair, the intensity of the 2$^{nd}$ harmonic is an order of magnitude smaller than the 1$^{st}$ harmonic. This intensity can only be enhanced by increasing the $I_{DC}$. However, change in $I_{DC}$ leads to variation in the spatial profile of the spin wave modes in the vicinity of the NCs (see supplementary information, Fig. S5) which may or may not favor the out-of-phase auto-oscillation[43]. Therefore, the phase binarization in our bias field-free NC SHNO pair is current tunable which is later utilized for implementation of RC scheme.

**Reservoir computing with phase-binarized SHNO pair**

We now demonstrate how the phase binarization can be utilized to perform efficient binary classification task using the NC SHNO pair as a reservoir. The learning task we choose is the classification of points that belong to randomly sequenced sin and square waveforms with identical period and amplitude. This is a standard RC benchmark task that requires significant nonlinearity as well as short-term memory. Notably, the points belonging to the extrema of sin and square waves are completely identical, therefore, the classification of these points are not trivial. One must remember the previous points to classify the extrema points. It has been observed that the phase binarization in the NC pair strongly depends on the recent history of input current in a continuous sweep of $I_{DC}$. This has been demonstrated in Fig. 5a and 5b. Here we observe the auto-oscillation in the full NC pair geometry at 2.3 mA input current pulse which follows an input current of 1 mA (Fig. 5a) and 1.5 mA (Fig. 5b) separately. We find that, in the first case, phase binarization is achieved at 2.3 mA as expected, which results in the reduced auto-oscillation amplitude in steady state (Fig. 5a). However, in stark contrast to the first case, the phase binarization is not achieved at 2.3 mA in the second case which results in distinctively higher auto-oscillation amplitude in steady state (Fig. 5b). Therefore, the phase binarization depends on the previous auto-oscillation state. The stable in-phase auto-oscillation state is reinforced by the spatial overlapping of the localized spin wave edge modes corresponding to the stronger $1^{st}$ harmonic. Therefore, it is difficult for the weaker $2^{nd}$ harmonic to induce phase binarization through SHIL. Hence, if the initial state is a steady in-phase auto-oscillation state, higher input current is required for phase binarization as compared to the stable initial state of $m$. We utilize this phenomenon as a short-term memory feature where the auto-oscillation state associated to the "present" input ($I_{DC}$ = 2.3 mA in Fig. 5a and 5b) is dictated by the "recent past" input ($I_{DC}$ = 1 mA or 1.5 mA in Fig. 5a and 5b). We further notice that the initial auto-oscillation state is restored once the input current is reduced back to 1 mA (Fig. 5a) or 1.5 mA (Fig. 5b), indicating the memory is not a long-term memory.

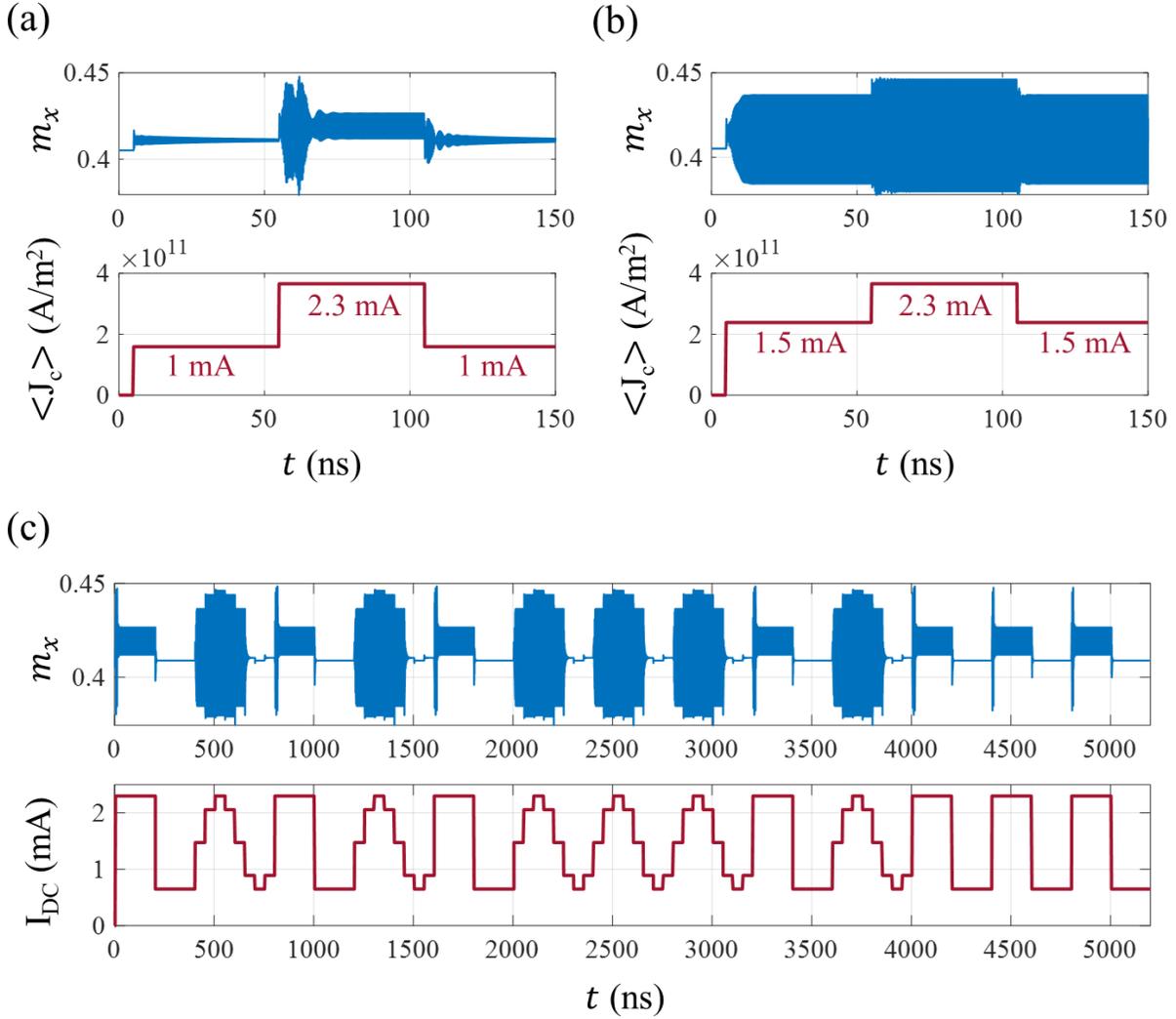

**Figure 5 | Towards sin and square wave classification with phase binarized NC pair.** The time evolution of $m_x$ in the NC SHNO pair ($d = 200$ nm) is shown for sequential excitation. (a) The time evolution of $m_x$ (top subfigure) corresponding to the $I_{DC}$ sequence 1mA→2.3mA→1mA (bottom subfigure). The auto-oscillation amplitude at 2.3 mA reduces at the steady state due to phase binarization. (b) Time evolution of $m_x$ for a slightly different $I_{DC}$ sequence: 1.5mA→2.3mA→1.5mA. Higher auto-oscillation amplitude at 2.3 mA at the steady state denotes in-phase auto-oscillation. The bottom subfigures in both (a) and (b) show the spatially averaged $J_c$ as function of time. (c) Temporal profile of $m_x$ for $I_{DC}$ sequence defined by randomly arranged sin and square waveforms.

Now we define a temporal sequence of randomly arranged sin and square waveforms of equal amplitude and period. Each waveform consists of eight points equally spaced in time. This input-time series is encoded in the magnitude of $I_{DC}$ as shown in Fig. 5c. For every single point in the time-series, the associated $I_{DC}$ is kept fixed for 50 ns to record the steady-state auto-oscillation profile. The time evolution of $m_x$ obtained from a continuous simulation for the entire sequence of $I_{DC}$ consisting of 104 input points (13 waveforms) is shown in Fig. 5c. We clearly observe the substantial difference in the auto-oscillation profiles corresponding to the sin and square waveforms. The phase binarization takes place only for points belong to the square waveform resulting in lower amplitude of $m_x(t)$. Therefore, the maxima points can be well classified despite the degeneracy in their values. However, the points in the lower half of the waveforms including the minima points do not produce any auto-oscillation as the $I_{DC}$ is lower than the threshold excitation current (~1.1 mA for $p = 200$ nm as seen from Fig. 1d). Therefore, these points cannot be classified using the present NC pair SHNO.

This issue could be resolved by adding two NiFe layers with uniaxial anisotropy, at both sides of the Pt layer (Fig. 6a). In this modified structure, the auto-oscillation can be induced in either one of the NiFe layer by altering the polarity of $I_{DC}$. As shown in Fig. 6a, the orientation of spin polarization and $\boldsymbol{m}$ in the bottom NiFe layer is such that the SOT opposes the damping torque. Hence, the bottom NiFe layer can exhibit auto-oscillation. On the other hand, in the top NiFe layer, the $\boldsymbol{m}$ and the spin polarization direction are oriented such a way that the SOT and the damping torque act parallel to each other. This leads to an enhancement of the effective damping in the FM layer and eventually relaxation of $\boldsymbol{m}$ to a stable equilibrium state. The scenario reverses for $I_{DC} < 0$, where only the top NiFe layer can exhibit auto-oscillation of $\boldsymbol{m}$. Therefore, the auto-oscillation can be obtained in such a trilayer SHNO irrespective of the polarity of input current.

We now reconstruct the $I_{DC}$ sequence in line with the main time-series such that the amplitudes of the sin and square waves span symmetrically about $I_{DC} = 0$. In addition, we modulate the $I_{DC}$ values through a standard time-multiplexing algorithm[44,45] (see supplementary note 7 for details). The time-multiplexing enables us to simulate multiple virtual neurons randomly connected to each other from a single dynamic node, which is the trilayer NC SHNO pair in our case. This essentially maps each point in the input to a higher dimensional space as per the requirement of the reservoir in RC. Here we simulate 20 virtual nodes (neurons). Therefore, each input point can be represented by the combined auto-oscillation state of all 20 nodes in the so called higher dimensional space.

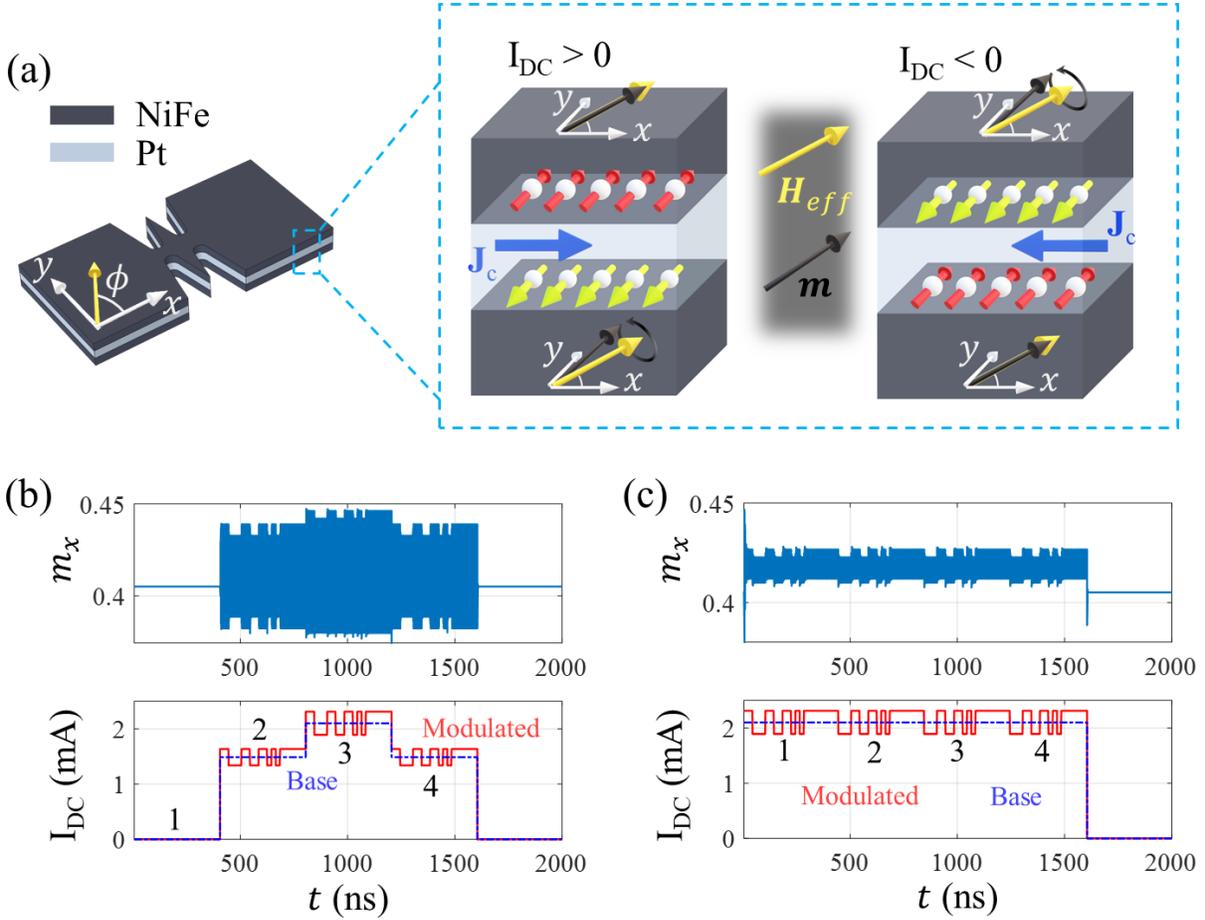

**Figure 6 | Modified trilayer SHNO geometry for RC**. (a) Schematic of the trilayer NC SHNO pair and the magnetization auto-oscillation under the change in polarity of $I_{DC}$. The yellow arrow in the trilayer NC SHNO pair geometry represents the easy axis that primarily decides the orintation of $H_{eff}$. (b) and (c) Time evolution of $m_x$ under the modulated $I_{DC}$ representing the time multiplexed inputs corresponding to the four consecutive points in the positive half of sin and square wave respectively. The "Base" represents the $I_{DC}$ values corresponding to the actual input data points. The time multiplexed i.e., pre-processed input for the RC is represented by the "Modulated" $I_{DC}$.

Figure 6b and 6c show the auto-oscillation profile for the positive half-cycles of sin and square waveform respectively, each consisting of four points. In Fig. 6b, the constant "base" $I_{DC}$ value representing a single point belonging to the sin waveform is shown in blue dash-dotted line. The corresponding "modulated" $I_{DC}$ values representing inputs for the virtual nodes are shown in continuous red line. Note that the oscillation amplitude fluctuates due to the modulation in $I_{DC}$ values, showing the randomness in the system. A similar auto-oscillation data obtained for the points in the positive half of the square-wave, are shown in Fig. 6c. We observe the noticeable difference in the auto-oscillation amplitude for sin and square wave points. The same auto-oscillation profiles have been obtained for the negative half of the waveforms due to auto-oscillation in the top NiFe layer. However, we still need to achieve a finite and relatively higher auto-oscillation amplitude for $I_{DC}$ = 0 mA, to classify the zero-value points in the sin wave category. This can be achieved by connecting an oscillator of comparable frequency and amplitude preceded by a NOT gate, in parallel to the NC pair (see supplementary information, Fig. S8).

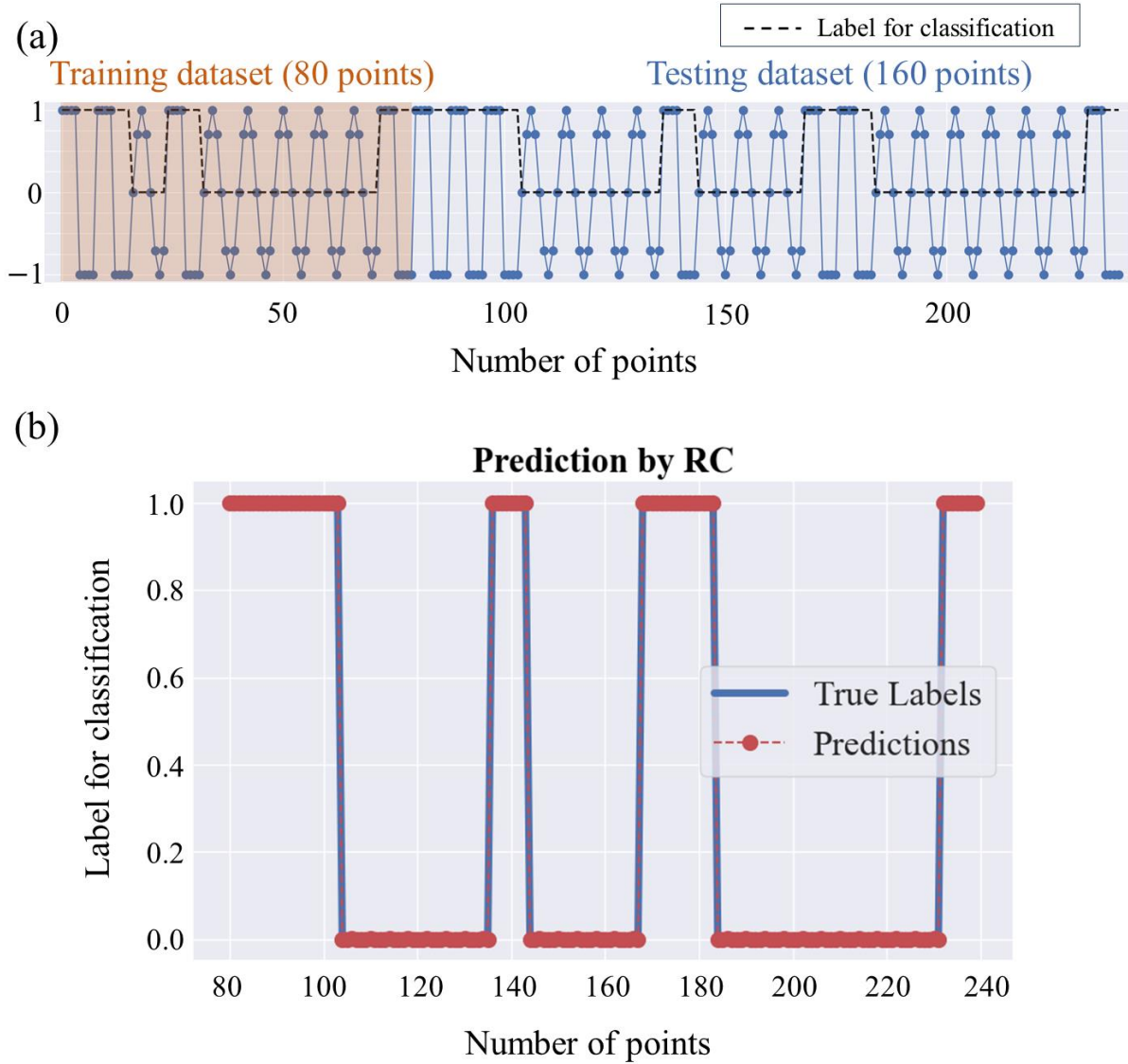

**Figure 7 | Performance of trilayer NC SHNO pair in sin and square wave classification task**. (a) The input points belonging to randomly sequenced sin and square waveforms. First 80 points have been used for training the reservoir. The RC performance in the classification task has been tested with the remaining 160 datapoints. The binary labels assigned to the input points are shown in dashed line. (b) Comparison between the true classifier labels of the points and the prediction by the NC SHNO pair as a reservoir. The match between the true labels and the predicted labels show 100% classification accuracy.

Finally, to demonstrate the classification performance of our NC pair SHNO as a reservoir, we define the time-series with 240 points (30 waveforms) as shown in Fig. 7a. We use the first 80 points for training and rest 160 points for testing. We further define the binary target labels to the points: "1" for the points in square wave and "0" for the points in sin wave (the black dotted line in Fig. 7a). Figure 7b shows the efficient prediction of the target labels by the reservoir which are 100% accurate with the true labels. The classification task has been performed with different random sequences of sin and square waves consisting of same periodicity and number of points. A few of the results have been shown in the supplementary information. Notably, in all these cases 100% accurate classification have been achieved.

## Discussion

We have demonstrated purely DC current-tunable mutual synchronization and phase binarization in a NC SHNO pair in absence of biasing magnetic field as well as external microwave for injection locking. The coupling between the two non-identical SHNOs is mainly mediated by the magnetodipolar interaction that leads to coherent synchronization of the SHNOs (see supplementary note 6). However, the characteristic frequencies of the single NC SHNOs are tens of MHz apart at lower input current. Therefore, mutual synchronization of such non-identical SHNOs is apparently difficult to achieve at low current regime (1.5 mA ≤ $I_{DC}$ ≤ 1.75 mA, see Fig. 2a). Nevertheless, both the NC SHNOs in our NC pair with $d$ = 200 nm, exhibit mutually synchronized auto-oscillation state in this regime. This has been possible through the modulation of the local anisotropy field at the NCs by the overlapping of the localized spin wave edge modes corresponding to the 1st harmonic. We recall that the local anisotropy field is directly determined by the instantaneous orientation of magnetization. The magnetodipolar coupling as well as overlapping of the localized spin wave edge modes regulate the orientation of instantaneous magnetization and reinforces coherent auto-oscillation. This, in turn modulates the local anisotropy field such a way that leads to similar instantaneous auto-oscillation frequencies in both SHNOs in the NC pair. In contrast to $d$ = 200 nm NC pair, there is no overlapping in the spin wave mode profile of the 1st harmonic in $d$ = 800 nm NC pair (Fig. 4d). Therefore, in that case, both the NCs exhibit auto-oscillation at their characteristic frequencies in unsynchronized state for low input current (see Fig. 4a). At higher values of input current, similar characteristic frequencies of both

NC SHNOs lead to mutual synchronization for both $d$ = 200 nm and $d$ = 800 nm NC pairs (Fig. 1d and 4a). In addition, we observe the phase binarization phenomena at mutually synchronized auto-oscillation state and its strong dependence on the recent history of dynamic magnetization. This phenomenon has been utilized for performing an RC benchmark classification task as demonstrated in this paper. It should be noted that, the RC scheme demonstrated here, is not to contrast different algorithm and propose a better alternative, but to highlight the physical phenomena of DC current-driven phase binarization in realizing a simple RC network.

While all the simulation results presented here are obtained without considering any thermal effect to focus more on understanding the physical origin and impact of the observed phenomena as well as to reduce simulation time, the effect of finite temperature (T = 300 K) has been presented in the supplementary information (Fig. S4). Similar phenomena of mutual synchronization and subsequent phase binarization have been observed for T = 300 K as well in the NC SHNO pair. Hence, our results hold promises to pave the way for overcoming challenges in synchronizing non-identical bias field-free NC SHNOs and utilize them for designing efficient RC hardware.

## Methods

**Device design in COMSOL**. The NC SHNO pair has been designed considering a bilayer stack of 5 nm thick NiFe layer interfaced with a 5 nm thick Pt layer. Both the round-edged NCs in the NC pair have been defined with 50 nm radius of curvature and 22° opening angle. The widths of the left and right NCs are defined as 100 nm and 150 nm (Fig. 1a). The centers of both NCs are colinear and equally apart from the longer sides of the device. The geometry of this bilayer NC SHNO pair has been designed in COMSOL Multiphysics software. In addition, the spatial distribution of $J_c$ in the Pt layer and the Oersted field in the NiFe layer corresponding to 1 mA input current ($I_{DC}$) have been simulated in COMSOL considering the standard conductivity values of 8.9 MS/m for Pt and 1.74 MS/m for NiFe.

**Micromagnetic simulation**. The micromagnetic simulations have been performed using the GPU-accelerated open source software MuMax3[40] that employs a finite difference discretization of space. A 2048 nm × 1024 nm × 5 nm simulation area has been uniformly discretized into

1024×512×1 rectangular grid. The relevant SHNO device geometry (NC Pair or single NC SHNOs) for micromagnetic simulation was extracted from COMSOL as a black and white image to define the NiFe region. MuMax3 solves the LLG equation (Eq. 2) in a ferromagnet. Therefore, only the NiFe layers of the SHNO devices have been simulated explicitly considering the Slonczewski-like spin-orbit torque term. The material parameters have been defined as follows[19,31]: $M_s$ = 600 kA/m, exchange stiffness constant, $A_{ex}$ = 10 pJ/m, intrinsic damping parameter, $\alpha$ = 0.02, gyromagnetic ratio, $\gamma$ = 29.53 GHz/T and uniaxial magnetic anisotropy constant, $K_u$ = 7.5 kJ/m³. To avoid the staircase effect at the circular edges of the NCs, the "edgesmooth" function of MuMax³ has been used. Finally, absorbing boundary condition[46] has been implemented to avoid the spurious reflection of spin wave from the boundaries.

The SOT term has been implemented in MuMax3 using the built-in Slonczewski spin-transfer torque (STT) model and disabling the Zhang-Li torque, following the approach of Dvornik et. al[19]. The spin Hall angle of Pt has been defined as $\theta_{SH}$ = 0.08[19,22,23] and assigned to the variable "Pol" in the Slonczewski STT model in MuMax³. The spin polarization direction ($\boldsymbol{\sigma} = -\hat{\boldsymbol{y}}$) has been implemented by setting the "Fixedlayer = vector (0, −1, 0)" command. In addition, the "epsilonprime" variable in the Slonczewski STT model has been set to "0" to neglect the field-like component of SOT in our simulation. To simulate only SOT driven magnetization dynamics, the initial magnetization has been relaxed into the ground state. Thereafter, a 5 ns initial delay has been introduced in the SOT to simulate the transient dynamics of magnetization in absence of any SOT and Oersted field and finally relax to a stable magnetization state. For each value of $I_{DC}$, the simulation has been carried out for 70 ns and the $m_x(t)$ of the last 30 ns (steady state auto-oscillation) has been analyzed. In case of T = 300 K, the simulation time has been fixed to 140 ns for each $I_{DC}$ value, while the $m_x(t)$ obtained from last 40 ns has been analyzed. The post-processing of the data has been done using custom built MATLAB and Python codes.

## Acknowledgement

S.M. acknowledges the support from the NTU-Research Scholarship (NTU-RSS). R.S.R. acknowledges the support from the National Research Foundation (NRF), Singapore through grant number NRF-CRP21-2018-0003 and the Ministry of Education, Singapore, through its Academic

Research Tier 1 grant number RG76/22. Any opinions, findings and conclusions or recommendations expressed in this material are those of the author(s) and do not reflect the views of the Ministry of Education, Singapore.

## Author Contribution

S. M. conceived the original idea. S. M. performed all the micromagnetic simulations. S. M. did all the analysis of the simulation results. S.M. drafted the manuscript. R.M., J.R.M., Y.F., and R. S. R. reviewed the manuscript and provided valuable inputs. R. S. R. supervised the project.

## Data Availability

The data that support the findings of this study are available from the corresponding author upon reasonable request.

## Code Availability

The MuMax3, Python and MATLAB codes used in this study are available from the corresponding author upon reasonable request.

## Competing interests

The authors declare no competing interests.

## Additional information

See the **Supplementary Information** for additional details.

# Supplementary Information

## Phase Binarization in Mutually Synchronized Bias Field-free Spin Hall Nano-oscillators for Reservoir Computing

Sourabh Manna[1], Rohit Medwal[2], John Rex Mohan[3], Yasuhiro Fukuma[3,4], Rajdeep Singh Rawat[1#]

[1]*Natural Sciences and Science Education, National Institute of Education, Nanyang Technological University, 637616, Singapore.*

[2]*Department of Physics, Indian Institute of Technology Kanpur, Uttar Pradesh, 208016, India.*

[3]*School of Computer Science and System Engineering, Kyushu Institute of Technology, Iizuka, 820-8502, Japan.*

[4]*Research Center for Neuromorphic AI hardware, Kyushu Institute of Technology, Kitakyushu 808-0916, Japan.*

# Correspondence: rajdeep.rawat@nie.edu.sg

1. Spatial profile of current density

Figure S1 below shows the spatial distribution of the x and y components of current density vector $J_c$ ($J_{cx}$ and $J_{cy}$ respectively) along with its magnitude $|J_c|$ in the NC pair with $d = 200$ nm. Comparing the $J_{cx}$ and $J_{cy}$ with $|J_c|$, it is evident that $J_{cx}$ is the dominant component of $J_c$. Therefore, $J_{cx}$ has been considered to formulate the SOT term in the simulation.

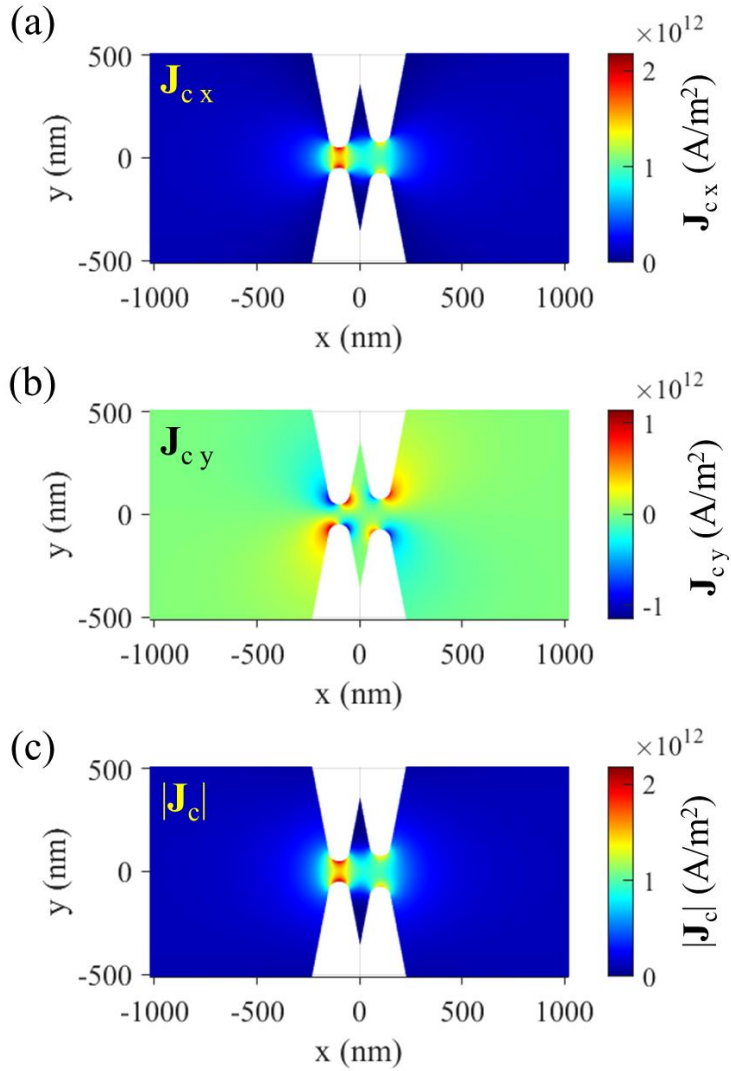

**Figure S1** | Spatial distribution profile of the x-component (a), y-component (b) of the current density vector ($J_c$) and (c) the magnitude of $J_c$ for $I_{DC} = 1$ mA.

## 2. Phase profile of the 2nd harmonic

Figure S2 shows the spatial profile of the phase of the 2nd harmonic of the magnetization auto-oscillation for (a) $I_{DC}$ = 1.5 mA and (b) $I_{DC}$ = 2.5 mA. The propagating nature of the associated spin wave mode is clear from the figure.

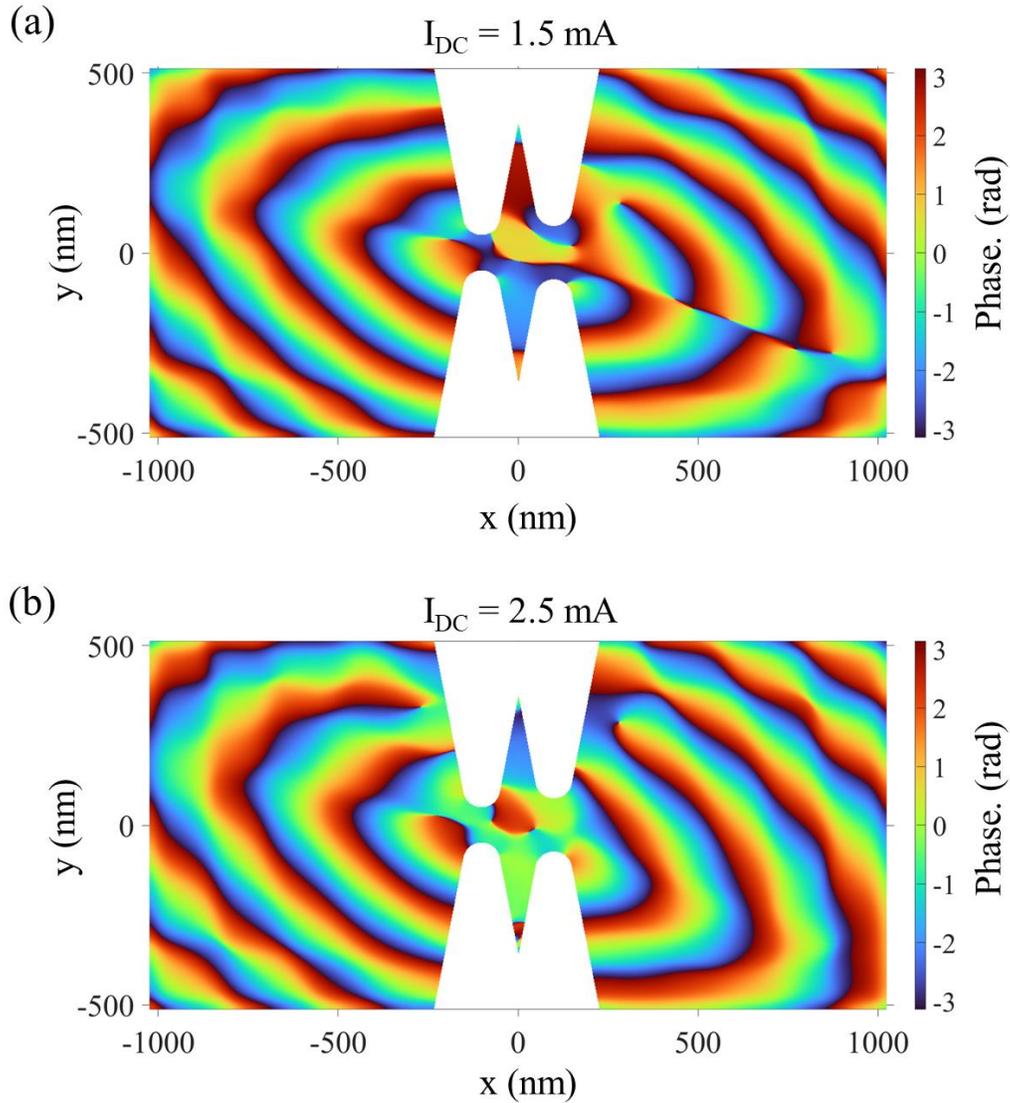

**Figure S2** | Spatial profile of the phase of 2nd harmonic for (a) $I_{DC}$ = 1.5 mA and (b) $I_{DC}$ = 2.5 mA showing the propagating nature of the spin wave mode.

## 3. Phase binarization for different values of $K_u$

Micromagnetic simulation has been carried out for different values of $K_u$ to observe the dependence of phase binarization phenomena on different uniaxial anisotropy field in the NC pair with $d = 200$ nm. Figure S3 presents the results. We observe that for $5$ kJ/m³ $\leq K_u \leq 8$ kJ/m³, the phase binarization threshold current shifts towards higher input current as $K_u$ increases. However, for $K_u = 9$ kJ/m³ and $10$ kJ/m³, the phase binarization threshold shifts towards lower input current. Notably, the best discretization has been achieved for $K_u = 7$ kJ/m3. The dependence of phase binarization on $K_u$ values can be explained considering the effect of demagnetization field that depends on the NC width. However, this is beyond the scope of this paper where we mainly focus on the mutual synchronization, phase binarization and their application in a ferromagnet with fixed uniaxial anisotropy.

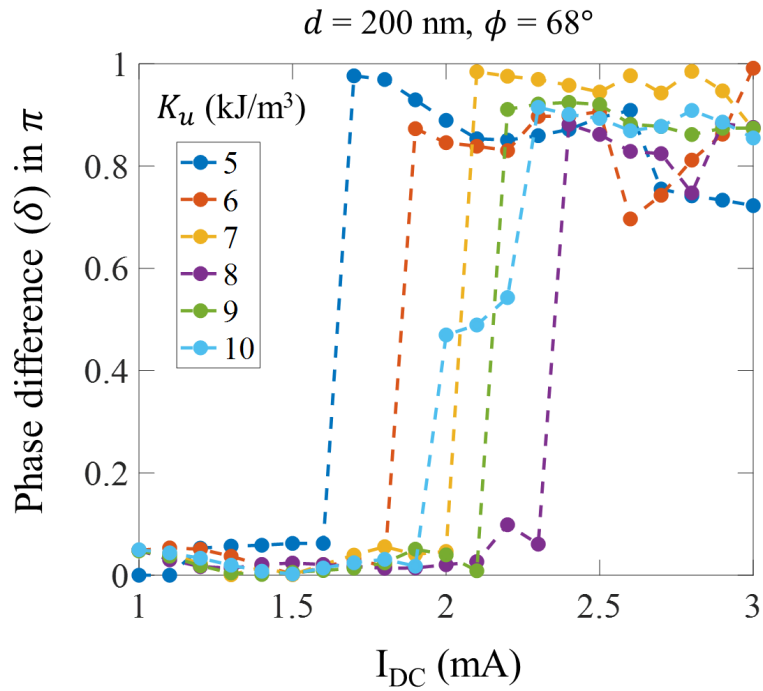

**Figure S3** | Phase difference between the NC SHNOs in the NC pair ($d = 200$ nm) as a function of $I_{DC}$ for different values of $K_u$.

## 4. Phase binarization at room temperature (T = 300 K)

Figure S4 depicts the auto-oscillation spectra at room temperature (T = 300 K), obtained from the NC pair with $d = 200$ nm (see Fig. 1a in the main text). The mutual synchronization and subsequent phase binarization have been observed in this case as well. The spectra exhibit a broader linewidth due to finite temperature. Notably, the phase binarization has been observed at lower threshold current as compared to T = 0 K case (see Fig. 1d in the main text). Therefore, we conclude that the random thermal fluctuation in the magnetization dynamics favors phase binarized state which results in lower threshold current for phase binarization. This can be useful for realizing the RC benchmark classification task of sin and square wave at lower input currents.

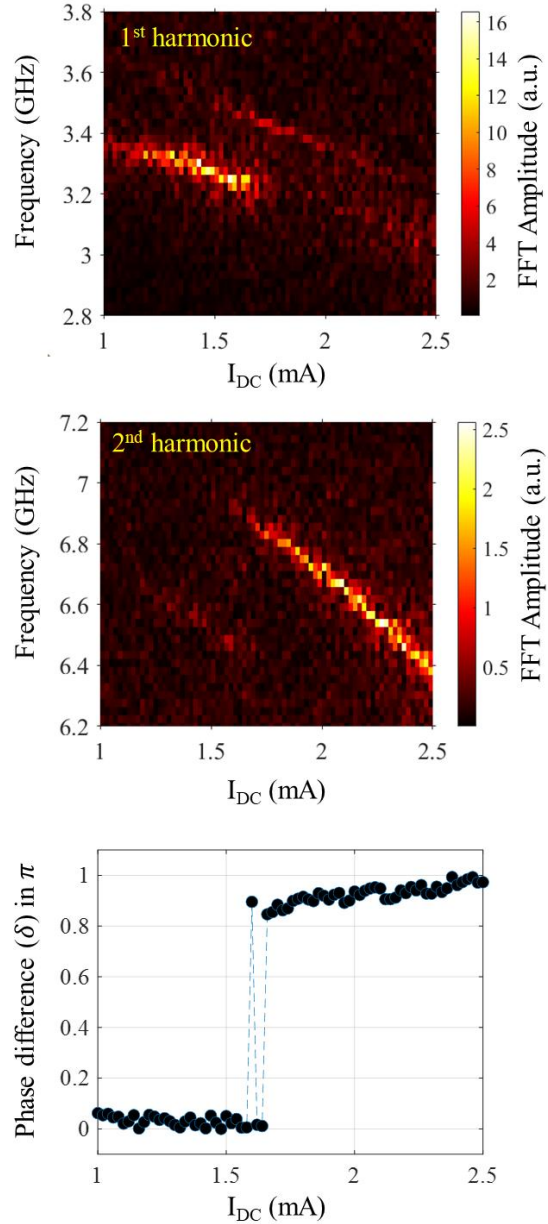

**Figure S4** | Auto-oscillation spectra obtained from the NC pair ($p$ = 200 nm) as function of $I_{DC}$ at (a) 1st harmonic and (b) 2nd harmonic at T = 300 K. The mutual synchronization and phase binarization have been observed here as well, denoting the consistency with the T = 0 case presented in the main text. (c) Phase difference ($\delta$) between the NC SHNOs in the NC pair as function of $I_{DC}$. The two discrete states in $\delta$ shows the phase binarization at higher values of $I_{DC}$.

## 5. Spatial profiles of the 1st harmonic at different values of $I_{DC}$

Here we show the spatial profiles of the spin wave modes and associated phase map of the 1st harmonic for NC pair with $d$ = 800 nm. As seen from the figure, the spin wave mode profile, especially for the propagating 2nd harmonic changes with $I_{DC}$ in the vicinity of the NC as well as in the bridge between the NCs. This variation in the spin wave mode profile may favor strong or weak phase binarization in the mutually synchronized auto-oscillation state. For example, weak phase binarization ($\delta \sim 0.4\pi$) is observed for $I_{DC}$ = 2.3 mA and 2.7 mA. However, strong phase binarization ($\delta \sim \pi$) is observed for $I_{DC}$ = 2.5 mA.

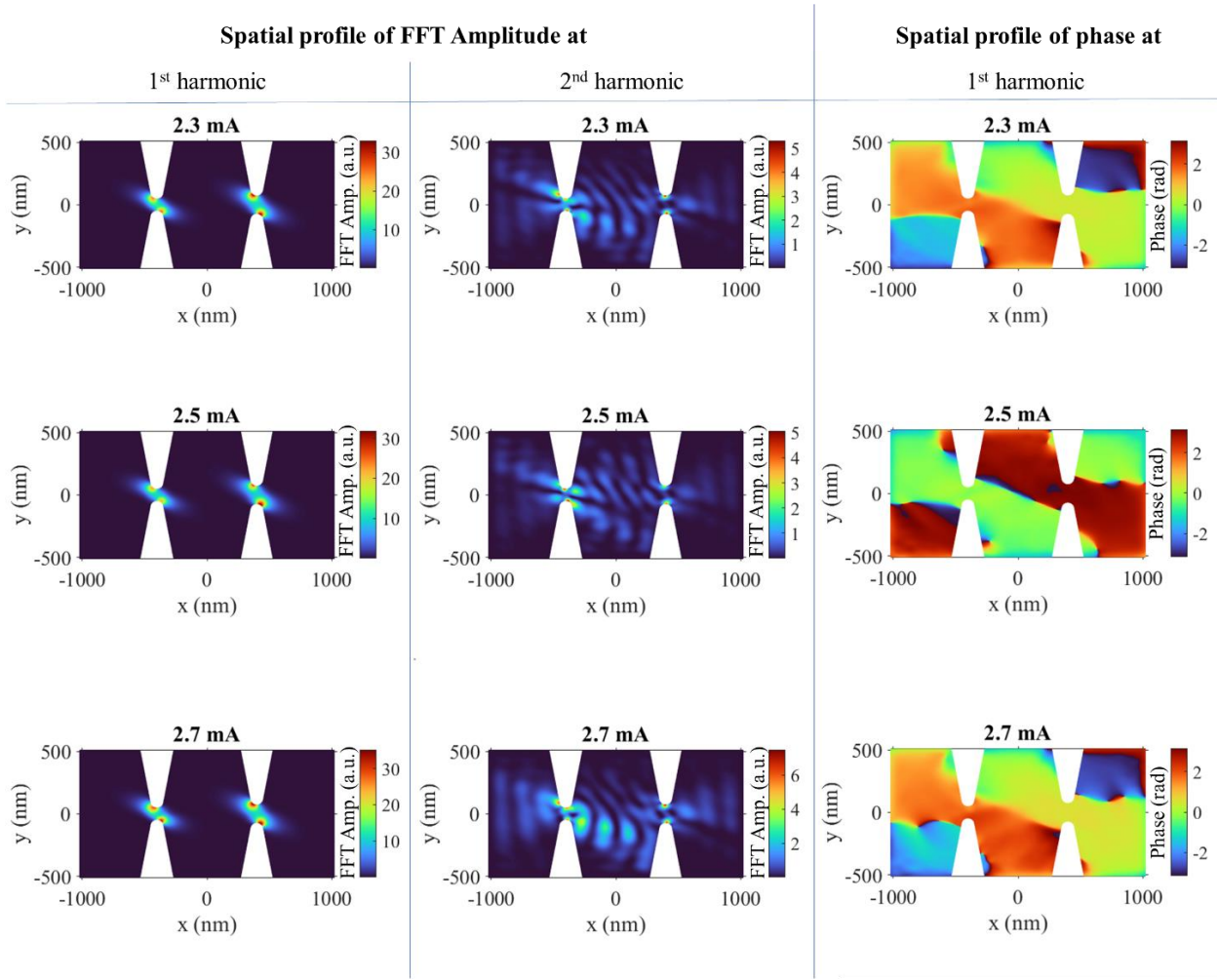

**Figure S5 |** Spatial profile of the spin wave modes corresponding to the 1st and 2nd harmonic along with the associated phase at 1st harmonic in the NC pair with $d$ = 800 nm.

## 6. Magnetodipolar coupling between the NC SHNOs in the NC pair.

The long-range magnetodipolar coupling is majorly responsible to mediate mutual synchronization of NC SHNOs in a chain or 2D pair of NCs[1-4]. We carried out micromagnetic simulations to understand the role of magnetodipolar coupling in our bias field-free NC SHNO pair. The simulation has been performed in the NC pair with $p$ = 800 nm to ensure that there is no overlapping of the localized spin wave edge modes corresponding to the 1$^{st}$ harmonic. Firstly, we defined a 20 nm wide discontinuity in the NiFe region along the y-axis to separate the two NCs as shown in Fig. S6a. This essentially creates two separate material regions which cannot interact by exchange interaction as the separation is much larger than the exchange length of NiFe ($\sim$ 5 nm). In addition, we explicitly turned off the exchange interaction between the left and right regions using "ext_ScaleExchange()" function in MuMax$^3$. We further employed absorbing boundary condition at the rectangular boundaries of both regions to stop the spin wave propagation between the two regions. Lastly, we deliberately turned off the current and SOT in the left region as shown in the inset of Fig. S6b. Therefore, the left NC should not exhibit any auto-oscillation of local magnetization. However, we have observed magnetization auto-oscillation locally at both left and right NCs. Figure S6b and S6c demonstrate the auto-oscillation characteristics. Notably, the auto-oscillation at the left NC is entirely driven by the auto-oscillation in the right NC as obvious from the exactly similar behavior of the FFT amplitude and frequency as function of $I_{DC}$ (in the right NC only) as observed from Fig S6b and S6c respectively. One can notice that the auto-oscillation frequency and the linewidth ($\Delta f$) are same for both NCs (Fig. S6c) that match with the characteristic behavior of auto-oscillation frequency of single NC SHNO with 150 nm NC width (see Fig. 2a). Hence, the local auto-oscillation in the left NC of 100 nm NC width is driven by the right NC. This also explains the smaller FFT amplitude of the auto-oscillation in the left NC (Fig. S6b). Since, there is no spin wave propagation or any exchange interaction between the NCs, they can interact only through the magnetodipolar coupling which induces magnetization auto-oscillation in the left NC in absence of any SOT.

In both NC pairs with $p$ = 200 nm and $p$ = 800 nm, the magnetodipolar interaction helps to establish mutual synchronization of the two NC SHNOs. However, the overlapping of the localized edge spin wave modes corresponding to the 1$^{st}$ harmonic locally modulates the anisotropy field to reinforce synchronization of the non-identical NC SHNOs. In addition, the spatial overlapping drives both the NC SHNOs towards the in-phase synchronization state.

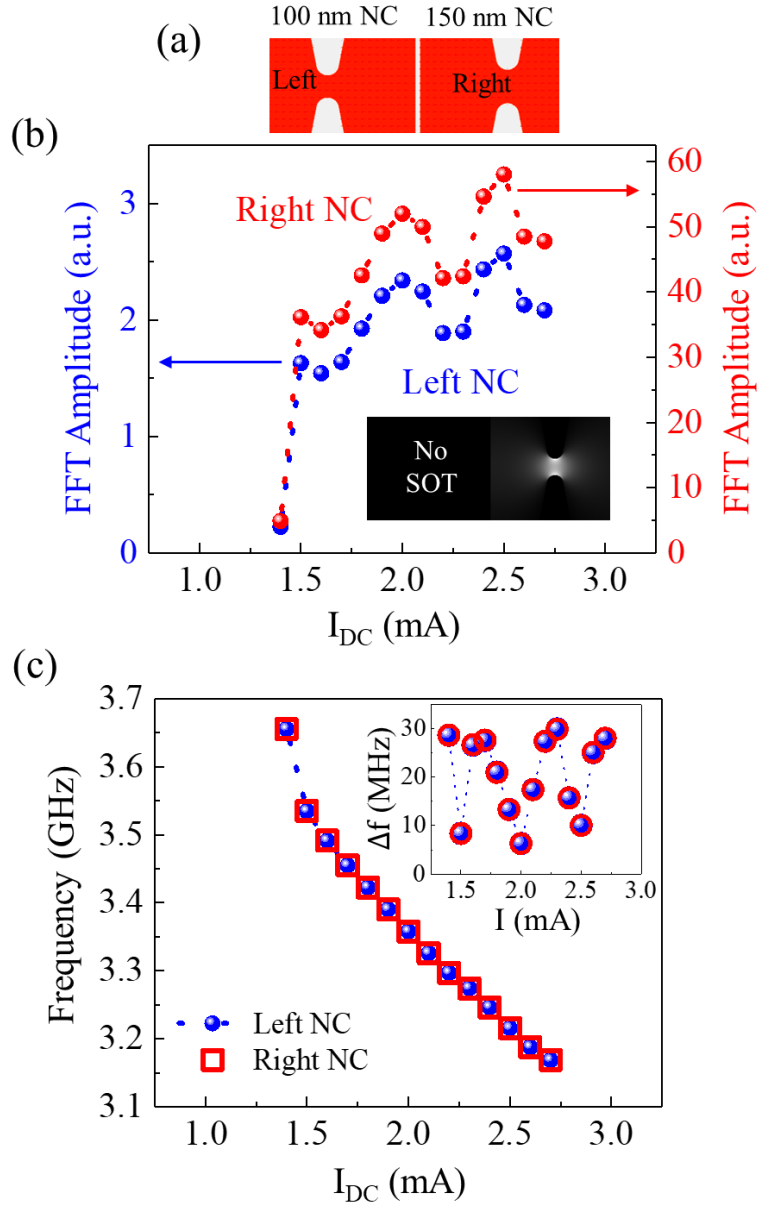

**Figure S6** | Testing for magnetodipolar coupling between the NCs. (a) NC pair ($p$ = 800 nm) with the 20 nm wide discontinuity in the material region. (b) FFT amplitude obtained from the local magnetization auto-oscillation at the two NCs as function of $I_{DC}$ considering no current and SOT in the left NC region (inset). (c) Variation of the auto-oscillation frequencies obtained from the local magnetization auto-oscillation at both NCs as function of $I_{DC}$. The variation of corresponding linewidth ($\Delta f$) is shown in the inset.

## 7. Preprocessing the input data for the classification task

Here we use the NC SHNO pair as a single dynamical node[5] for performing the RC benchmark classification task: sin and square wave classification. The next step is, therefore, to emulate virtual temporal neurons for constructing the reservoir which is a randomly interconnected recurrent neural network. This is implemented through a standard time-multiplexing of the input data as shown in the Figure S7. We multiplied each value (or point) in the input data sequence by a random binary vector with 20 elements, each of value $+1$ or $-1$. This masking of the input data generates the "pre-processed inputs". This process has been shown for four consecutive points in the sin wave (Fig. S7a) and square wave (Fig. S7b). Hence, each point in the input sequence is represented by the sequence of 20 points in the pre-processed input. The absolute values of these 20 points are same as the absolute value of the parent input point, as can be seen by comparing the pre-processed input with the input data in Fig. S7. Each point of this 20-point sequence act as the input to an individual node (neuron) in the reservoir. This way, one single point of input data is sent to all the nodes in the reservoir with random interconnection among themselves. Notably this random connection remains fixed for all the points in the input sequence as the random interconnection between the nodes in the reservoir should be time-invariant. Finally, we encode the pre-processed input into $I_{DC}$ by modulating its magnitude as shown in the "Input current" subfigures in Fig. S7. The "Base" $I_{DC}$ denotes the primary $I_{DC}$ value corresponding to the individual points in the input data whereas, the "Modulated" $I_{DC}$ denote the encoding of pre-processed input. The temporal evolution of $m_x$ corresponding to this modulated input current has been shown in Fig. 6b and 6c in the main text.

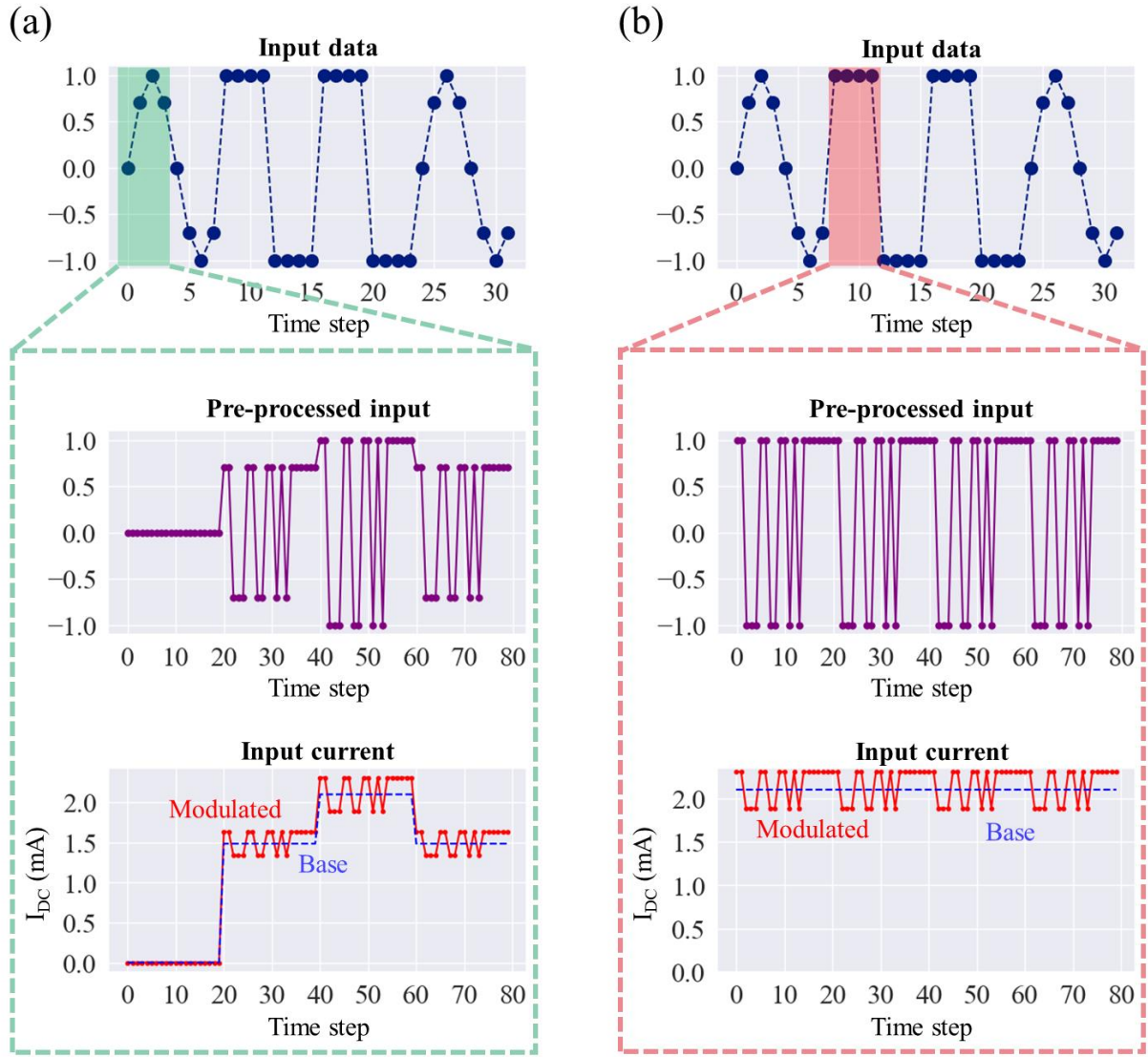

**Figure S7** | Steps for encoding the main input data into the modulated $I_{DC}$ as the reservoir input, shown for (a) sin wave and (b) square wave. Similar steps are followed for the negative half of both waveforms.

8. **Schematic of the proposed experimental arrangement for classification with trilayer NC SHNO pair.**

Figure S8 illustrates the experimental setup for classifying sine and square waves. The NOT gate triggers the electronic oscillator when $I_{DC} = 0$, generating an oscillatory voltage signal with an amplitude matching the auto-oscillation amplitude at the input points in the sine waveform.

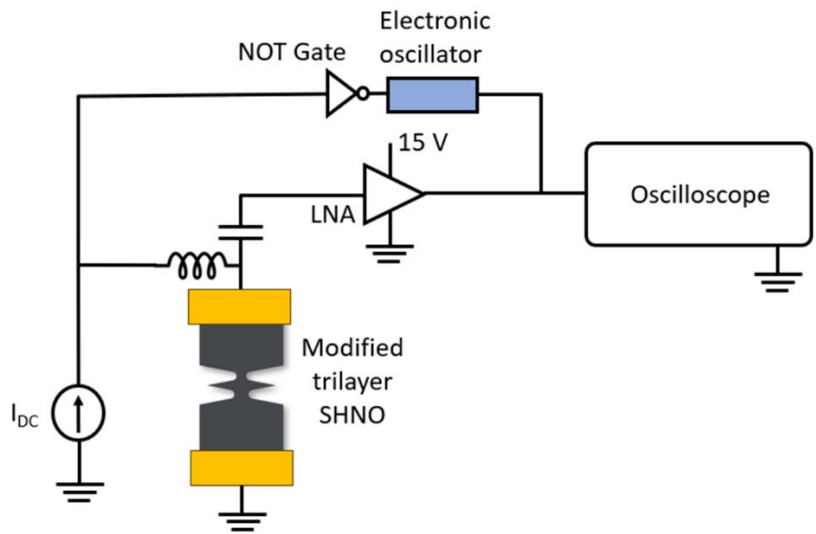

**Figure S8 |** Schematic of a proposed circuit to implement the RC classification task with trilayer NC SHNO pair. The $I_{DC}$-source generates the modulated $I_{DC}$ values as the input of the reservoir.

9. **Sin and Square wave classification for six different sequences of sin and square waveforms**

Figure S9 below shows the classification performance of the NC pair for six different sequences of sin and square waveforms. We observe the 100% classification accuracy for all the sequences.

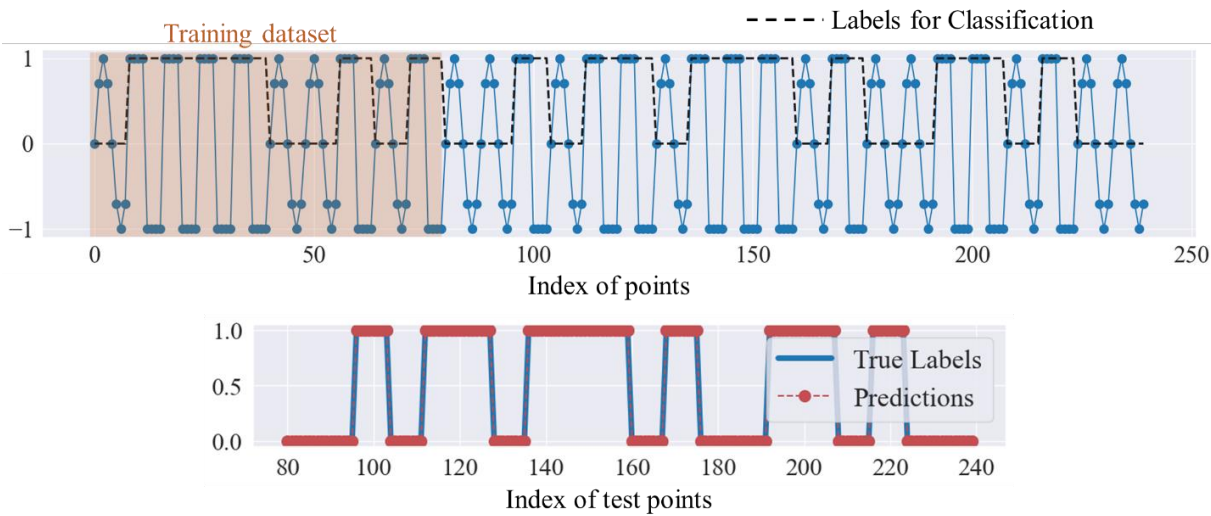

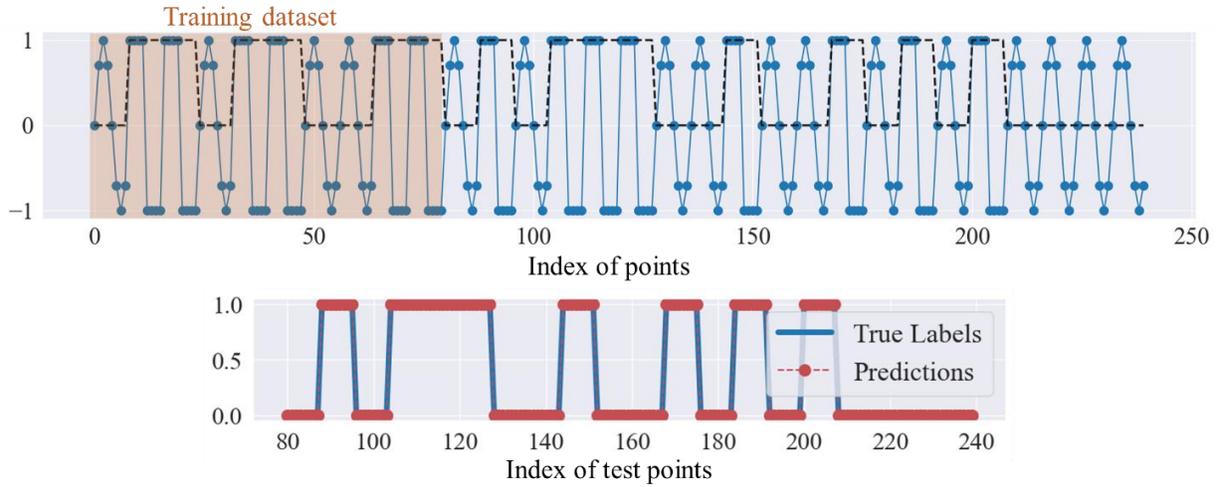
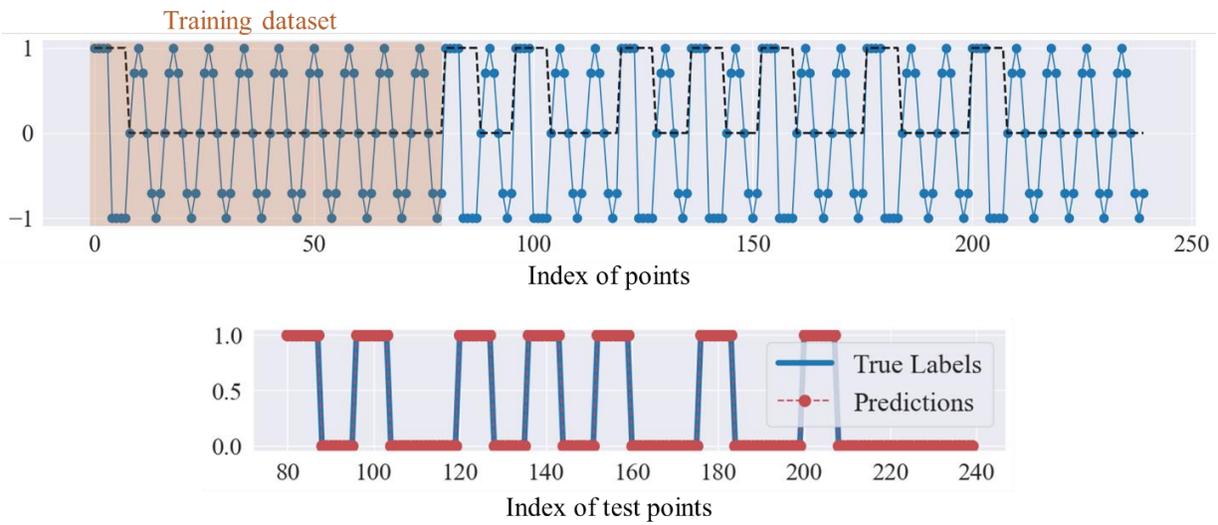
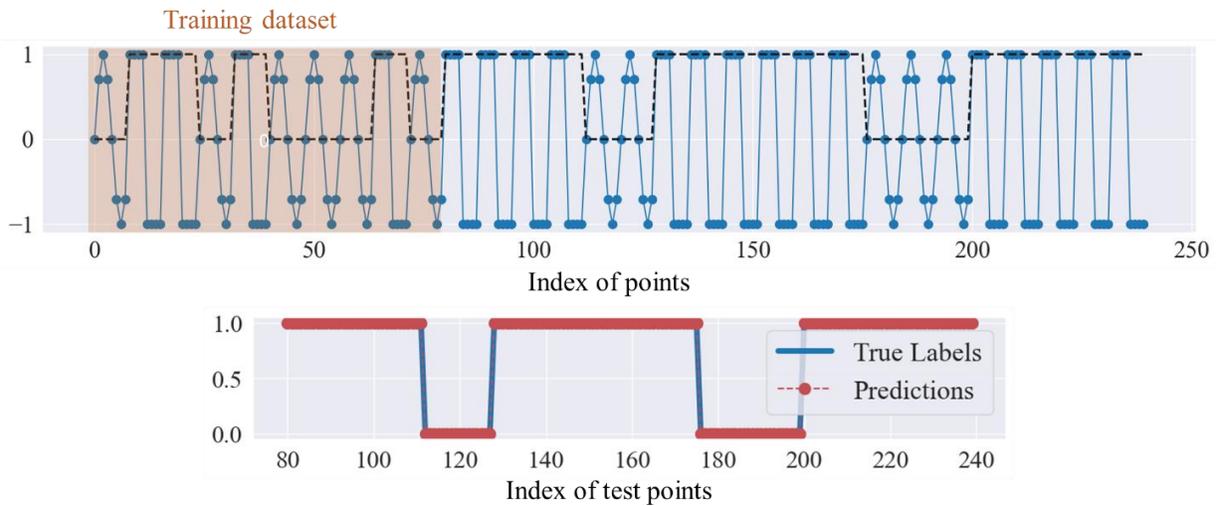

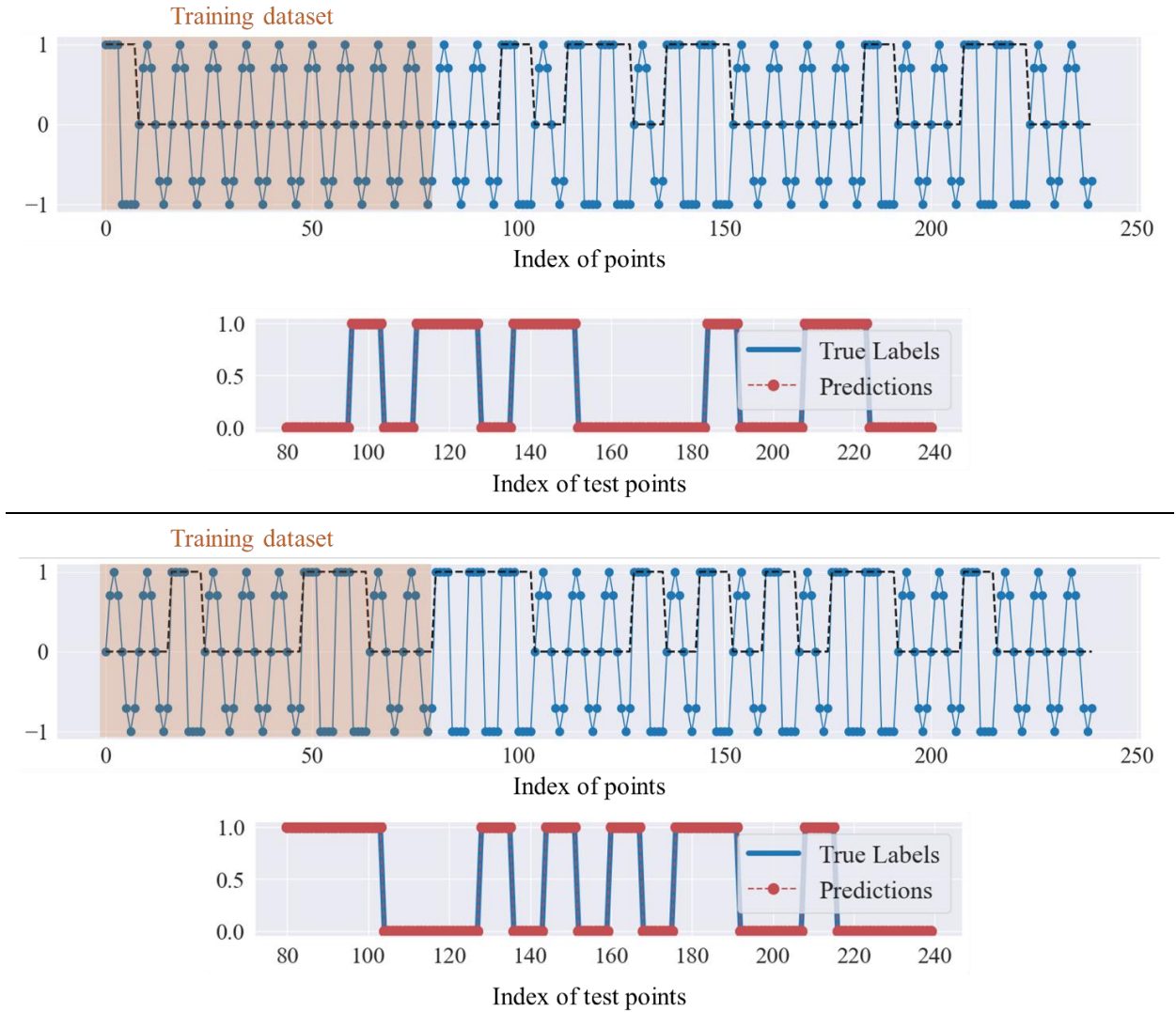

**Figure S9** | Examples of the sin and square wave classification task performed by the modified NC SHNO pair ($d =$ 200 nm) for six different sequences of sin and square waveforms. The black dashed black line shows the assigned binary labels for classifications. The matching between the true labels of the testing data and the labels predicted by the RC shows 100% accuracy in the classification.